\documentclass[showpacs,superscriptaddress,nofootinbib,floatfix,11pt]{revtex4-2}
\usepackage{setspace}
\usepackage[utf8]{inputenc}
\usepackage{float}
\usepackage{amsmath,amssymb,amsfonts,bm}
\usepackage{graphicx}
\usepackage{multirow}
\usepackage[usenames,dvipsnames]{color}
\allowdisplaybreaks 
\usepackage[
colorlinks=true,
linkcolor=blue,
breaklinks=true,
citecolor=blue]{hyperref}
\graphicspath{{figs.dir/}}

\newcommand{\be}{\begin{equation}} 
\newcommand{\ee}{\end{equation}}
\newcommand{\bea}{\begin{eqnarray}} 
\newcommand{\eea}{\end{eqnarray}}
\newcommand{\beas}{\begin{eqnarray*}} 
\newcommand{\eeas}{\end{eqnarray*}}

\newcommand{\lag}{\mathcal{L}}

\newcommand{\mev}{\text{MeV}}

\newcommand{\zc}{Z_c(3900)}
\newcommand{\zcpm}{Z_c (3900)^{\pm}}
\newcommand{\zcneu}{Z_c(3900)^0}
\newcommand{\zcs}{Z_{cs}(3985)}
\newcommand{\zcsmi}{Z_{cs}(3985)^-}
\newcommand{\jp}{J/\psi}
\newcommand{\jpi}{J/\psi \pi^{\pm}}
\newcommand{\Ds}{D^{(*)}}
\newcommand{\barDs}{\bar{D}^{(*)}}
\newcommand{\A}{\mathcal{A}}
\newcommand{\C}{\mathcal{C}}

\newcommand{\itp}{\affiliation{CAS Key Laboratory of Theoretical Physics, Institute of Theoretical Physics,\\ Chinese Academy of Sciences, Beijing 100190, China}}

\newcommand{\ucas}{\affiliation{School of Physical Sciences, University of Chinese Academy of Sciences, Beijing 100049, China}}

\newcommand{\ific}{\affiliation{Instituto de F\'isica Corpuscular (centro mixto CSIC-UV),
Institutos de Investigaci\'on de Paterna, \\ Apartado 22085, 46071 Valencia, Spain
}}

\begin{document}

\title{\boldmath A combined analysis of the $Z_c(3900)$ and the $Z_{cs}(3985)$ exotic states}

\author{Meng-Lin Du}\email{du.menglin@ific.uv.es}
\ific

\author{Miguel Albaladejo}\email{Miguel.Albaladejo@ific.uv.es}
\ific

\author{Feng-Kun Guo}\email{fkguo@itp.ac.cn}
\itp
\ucas

\author{Juan Nieves}\email{jmnieves@ific.uv.es}
\ific


\begin{abstract} 
  We have performed a combined analysis of the BESIII data for both the $\zc$ and $\zcs$ structures, assuming that the latter is an SU(3) flavor partner of the former one. We have improved on the previous analysis of Ref.~\cite{Albaladejo:2015lob} by computing the amplitude for the $D_1\bar{D}D^*$ triangle diagram  considering  both  $D$ and $S$-wave $D_1D^*\pi$ couplings. We have also investigated effects from SU(3) light-flavor violations, which are found to be moderate and of the order of 20\%. 
The successful reproduction of the BESIII spectra, in both the hidden-charm and hidden-charm strange sectors, strongly supports that the $\zcs$ and $\zc$ are SU(3) flavor partners placed in the same octet multiplet. The best results are obtained when an energy-dependent term in the diagonal $D^{(*)}\bar D_{(s)}^{(*)}$ interaction is included, leading  to resonances (poles above the thresholds) to describe these exotic states. We have also made predictions for the isovector $Z_{c}^*$ and isodoublet $Z_{cs}^*$, $D^*\bar{D}^*$ and $D^*\bar{D}_{s}^*$ molecules, with $J^{PC}=1^{+-}$ and $J^{P}=1^{+}$, respectively. These states would be heavy-quark spin symmetry (HQSS) partners of the $Z_{c}$ and $Z_{cs}$. Besides the determination of the masses and widths of the  $\zc$ and $\zcs$, we also predict those of the $Z_c^*$ and $Z_{cs}^*$ resonances. 

\end{abstract}

\maketitle

\section{Introduction}

The discovery of the $\chi_{c1}(3872)$~\cite{Belle:2003nnu}, also known as $X(3872)$, in 2003 built a landmark in the study of strong interactions and opened the gate to the abundance of the $XYZ$ structures in the heavy quarkonium region. Many of them are difficult to be understood from the conventional quark model point of view and thus turn out to be excellent candidates for exotic states, {\it i.e.}, hadrons with a content other than a quark-antiquark pair ($q\bar{q}$) or three quarks ($qqq$). 
A large amount of experimental and theoretical studies are devoted to those $XYZ$ states, see {\it e.g.} Refs.~\cite{Cincioglu:2016fkm,Chen:2016spr,Chen:2016qju,Esposito:2016noz,Guo:2017jvc,Olsen:2017bmm,Liu:2019zoy,Brambilla:2019esw,Guo:2019twa,Albaladejo:2020tzt,JPAC:2021rxu}. Among these states, the charged $\zc$~\cite{BESIII:2013ris,Belle:2013yex} and $\zcs$~\cite{BESIII:2020qkh} states are of particular interest since they are close to the $D\bar{D}^*$ and $D^*\bar D_s$/$D\bar{D}_s^*$ thresholds, respectively, with widths of the same order, which suggests that the latter be a candidate to be the strange partner of the former.

The structure associated to the $\zcpm$ was simultaneously first observed in the $\jpi$ mass spectrum of the  $e^+e^-\to \jp \pi^+\pi^-$ reaction  by the BESIII~\cite{BESIII:2013ris}  and Belle~\cite{Belle:2013yex} collaborations. In BESIII,  the $e^+e^-$ center of mass (c.m.) energy was fixed to $M=4.26$~GeV, while in the Belle experiment, using the initial state radiation method, the energy was  taken in the $Y(4260)$ region.  The peak was confirmed in an analysis of the CLEO-c data of the $\jpi$ mass spectrum, with a slightly lower c.m. energy of the $e^+e^-$ pair, at about $4.17$~GeV~\cite{Xiao:2013iha}. 
Further experimental evidence for the $\zcpm\to J/\psi\pi^\pm$ came from the semi-inclusive decays of $b$-flavored hadrons in the D0 experiment~\cite{D0:2018wyb}, with the $J/\psi\pi^+\pi^-$ invariant mass also constrained around the $Y(4260)$ mass region.

In addition, the first evidence of the neutral $\zcneu$ decaying into $\jp \pi^0$, was reported in Ref.~\cite{Xiao:2013iha} using the CLEO-c data. Later, a similar charged structure, named as $Z_c(3885)^+$, was observed in the $(D\bar{D}^*)^+$ mass distribution by the BESIII Collaboration~\cite{BESIII:2013qmu,BESIII:2015pqw}. The angular analysis showed that the favored spin-parity  quantum numbers for this peak were $J^P=1^+$. The similarity of the masses and widths of the $Z_c(3900)^+$ and $Z_c(3885)^+$ peaks  suggests a common origin for both of them. 

Extensive theoretical research has been done to understand the nature of the $\zc$ structure, providing explanations for it   as a  compact tetraquark~\cite{Braaten:2013boa,Dias:2013xfa,Maiani:2014aja,Qiao:2013raa,Deng:2014gqa} or as a $D\bar D^*$ resonant or virtual molecular state~\cite{Wang:2013cya,Wilbring:2013cha,Guo:2013sya,Dong:2013iqa,Zhang:2013aoa,Aceti:2014uea,Albaladejo:2015lob,Albaladejo:2016jsg,Du:2020vwb,Wang:2020dgr}. The possibility that the observed peaks had a kinematic origin was also discussed in Refs.~\cite{Chen:2013coa,Swanson:2014tra,HALQCD:2016ofq,Pilloni:2016obd}. However, it was demonstrated in Ref.~\cite{Guo:2014iya} that a pure kinematic two-body threshold cusp, without a near-threshold pole, is unlikely to produce a narrow peak in the mass distribution of the elastic channel of the pair of open heavy-flavor mesons, which has the threshold in the vicinity of the observed peak.\footnote{In the lattice simulation of the HAL QCD Collaboration, performed with pion masses ranging from 410 to 700~MeV, a pole more than 100~MeV (with a large uncertainty) below the $D\bar D^*$ threshold was found in the $J/\psi\pi$-$\eta_c\rho$-$D\bar D^*$ coupled-channel space~\cite{HALQCD:2016ofq,Ikeda:2017mee}. However, the predicted $D\bar D^*$ distribution is much broader than the experimental one reported by BESIII~\cite{BESIII:2015pqw}.} 
Furthermore, it was shown in Ref.~\cite{Dong:2020hxe} that the half-maximum width of the two-body threshold cusp lineshape is proportional\footnote{This is the case when the interaction between the hadron pair is attractive, but it is not strong enough to form a bound state. In this case, the pole is a virtual state, and the maximum of the absolute value of the amplitude is exactly at the threshold, where a cusp is produced (see also Refs.~\cite{Guo:2017jvc,Brambilla:2019esw}).} to $1/(\mu a_0^2)$, where $\mu$ and $a_0$  are the reduced mass and  the $S$-wave scattering length of the hadron pair, respectively. Therefore, a pronounced peak would imply a large scattering length, unless the reduced mass is very big, and consequently it points out to the existence of a near-threshold pole.
Thus, the pure two-body threshold cusp effect interpretation of the $XYZ$ states, which show up as pronounced near-threshold peaks, is not favored, and unitarity (or re-summing up the $s$-channel loops) will necessarily lead to a near-threshold pole with such an interaction. 

However, the situation becomes more complicated when there can be triangle singularities  in the same near-threshold region. 
A triangle singularity (TS) is a logarithmic branch point~\cite{Landau:1959fi,Coleman:1965xm,Bayar:2016ftu}, which could produce a peak resembling a resonance when it is close to the physical region (for a detailed review, we refer to Ref.~\cite{Guo:2019twa}).
The manifestation of the TS in the $D_1(2420)\bar D D^*$ loop in the $\jpi$ distribution, relevant to the $\zc$ structure in the reaction $e^+e^-\to \jp \pi^+\pi^-$, was immediately noticed~\cite{Wang:2013cya,Wang:2013hga} after the $\zc$ discovery.
It was demonstrated that the triangle diagrams need to be properly taken into account to reproduce the spectrum, while they alone can not satisfactorily describe the $\jpi$ mass distribution. In Ref.~\cite{Albaladejo:2015lob}, the final-state interactions for the $J/\psi\pi$-$D\bar D^*$ coupled channels are accounted for in the re-scattering $T$-matrix, calculated using the proper triangle diagrams. 
Such a scheme provides a satisfactory description of the $e^+e^-\to Y(4260)\to \jp \pi^+\pi^-$ and $Y(4260)\to \pi^{+}(D^0D^{*-})$ data simultaneously. Two different scnearios are considered to describe the $D\bar{D}^*$ interaction, and in each case only one pole is found to be identified as the $Z_c$ state~\cite{Albaladejo:2015lob}, which implies that the $Z_c(3900)^+$~\cite{BESIII:2013ris,Belle:2013yex} and $Z_c(3885)^+$~\cite{BESIII:2013qmu,BESIII:2015pqw} have the same origin and are generated by the $D\bar{D}^*$ interaction. 
On the other hand, the JPAC collaboration concluded that the data could also be described if there is no pole in the near-threshold region~\cite{Pilloni:2016obd}.\footnote{In the fit corresponding to the ``no pole" scenario (denoted as ``tr.''),  the JPAC analysis~\cite{Pilloni:2016obd} includes a penalty in the merit $\chi^2$ function to exclude a near-threshold pole. The authors found that this scenario cannot be statistically rejected. We nevertheless point out that even in this case, the amplitude would still allow for the existence of a pole, though its position, being far from threshold, was not given in the JPAC paper.}
A difference between the analyses carried out in Refs.~\cite{Albaladejo:2015lob} and \cite{Pilloni:2016obd} is that a $D$-wave $D_1D^*\pi$ vertex was considered in the former, while an $S$-wave  one was assumed in the latter.

Later on, new charged charmoniumlike structures $Z_c(4025)^\pm $ were observed near the $(D^*\bar{D}^*)^\pm$ threshold in the $\pi^\mp$ recoil mass spectrum~\cite{BESIII:2013mhi} and another charged peak, $Z_c(4020)^\pm$, was reported in the $h_c\pi^\pm$ invariant mass distribution~\cite{BESIII:2013ouc}. While the mass of $Z_c(4020)$ is close to that of the $Z_c(4025)$, the reported width of the $Z_c(4020)$ is larger than that of the $Z_c(4025)$, although the resonance parameters of both exotic states  agree  within 1.5\,$\sigma$~\cite{BESIII:2013ouc}. The proximity of the $Z_c(4020)$/$Z_c(4025)$ to the $D^*\bar{D}^*$ threshold suggests that it could be a proper candidate for the heavy quark spin symmetry (HQSS) partner of the $Z_c(3900)$, which would imply that the spin-parity of this isovector resonance would be $J^P=1^+$ ~\cite{Guo:2013sya,Nieves:2012tt,Hidalgo-Duque:2012rqv}. 

The BESIII collaboration reported in Ref.~\cite{BESIII:2020qkh} the first signal of  a hidden-charm resonant structure with strangeness, $\zcsmi$, observed in the $K^+$ recoil-mass spectrum of the reaction $e^+e^-\to K^+(D^{*0}D_s^-+D^0D_s^{*-})$ for events collected at a c.m. energy $\sqrt{s}=4.681$ GeV for the $e^+e^-$ pair. The pole mass and width were determined as $(3982.5^{+1.8}_{-2.6}\pm 2.1)~\mev$ and $(12.8^{+5.3}_{-4.4}\pm 3.0)~\mev$, respectively. The proximity of $\zcs$ to the $D^{*0}\bar{D}_s$  and $D^0\bar{D}_s^*$ thresholds, which are located at 3975~MeV and 3977~MeV, respectively, immediately spurred the hadronic molecular interpretation of this state~\cite{Yang:2020nrt,Meng:2020ihj,Sun:2020hjw,Wang:2020htx,Xu:2020evn,Wang:2020dgr,Yan:2021tcp,Ortega:2021enc,Wu:2021ezz,Baru:2021ddn}, as a strange partner of the $\zc$. 
Other possible interpretations for the nature of the $\zcs$ were also suggested, see {\it e.g.} Refs.~\cite{Wang:2020kej,Wan:2020oxt,Chen:2020yvq,Wang:2020rcx,Wang:2020iqt,Jin:2020yjn,Ikeno:2020mra,Guo:2020vmu,Karliner:2021qok}. An analysis of the $D^{*0}\bar{D}_s+D^0\bar{D}_s^*$ invariant mass spectra was performed in Ref.~\cite{Yang:2020nrt}, in which the  $D^*\bar{D}_s$ and $D\bar{D}_s^*$ $S$-wave interactions ($J^{P}=1^{+}$) were related to that of the  $D\bar{D}^*$ pair in the  $J^{PC}=1^{+-}$ channel using the SU(3) light-quark flavour symmetry. There, in addition to the direct point-like production of $K^+D^{*0}\bar{D}_s$ and $K^+ D^0\bar{D}_s^*$, a triangle diagram mechanism with the loop $D_{s2}^*\bar D_s^{*}D^0$ was also considered  because of the presence of a nearby TS, which can mimic the peak structure and enhance the production of near threshold molecules~\cite{Guo:2019twa,Guo:2020oqk}. 
The analysis was, in principle, improved in Ref.~\cite{Baru:2021ddn} by incorporating the $\bar{D}_s^*D^*/\bar{D}^*D_s^*$ channels and extending the analysis to the whole energy range covered by the BESIII data. Two types of solutions that describe the data almost equally well are found in Ref.~\cite{Baru:2021ddn}, and both are consistent with the interpretation of the $\zcs$ sate as an SU(3) partner of the $\zc$. In Ref.~\cite{LHCb:2021uow}, the LHCb collaboration reported, from an  analysis of the $B^+\to \jp \phi K^+$ decay amplitude, two new hidden-charm strange structures $Z_{cs}(4000)^+$ and $Z_{cs}(4220)^+$ decaying into $J/\psi K^+$. 
In particular, the $Z_{cs}(4000)^+$ was determined to have a mass of $(4003\pm 6^{+\phantom{1}4}_{-14})$ MeV, a width of $(131\pm 15\pm26)$ MeV and  $J^P=1^+$ spin-parity. While its mass is close to that of the $Z_{cs}(3985)$, its width is much larger than that of the $Z_{cs}(3985)$, which has generated a debate on whether the $\zcs$ and the $Z_{cs}(4000)$ correspond to the same state~\cite{Yang:2020nrt,Ortega:2021enc} or not~\cite{Maiani:2021tri,Meng:2021rdg} (see also Ref.~\cite{Ikeno:2021mcb}). 

In Ref.~\cite{Yang:2020nrt}, the SU(3) strangeness  partner of the $\zc$ was predicted from the pole position obtained for this state in the study of Ref.~\cite{Albaladejo:2015lob}. In that work, fits to the $K^+$ recoil-mas distributions of the process $e^+e^-\to K^+(D^{*0}\bar{D}_s+D^0\bar{D}_s^*)$, for events collected at different energy points by BESIII, were also carried out (see Figs. 3 and 4 of Ref.~\cite{Yang:2020nrt}).  A difference of a few tens of MeV was found between the two mass determinations, which is largely covered by the uncertainties. 

In the present work, we perform a combined analysis of the BESIII data for both the $\zc$ and $\zcs$ structures assuming that the $\zcs$ is an SU(3) flavor partner of the $\zc$. Additionally, we improve on a different aspect.  In the analysis of Ref.~\cite{Albaladejo:2015lob}, the amplitude for the $D_1\bar{D}D^*$ triangle diagram  considered a $D_1D^*\pi$ coupling in $D$-wave. However,  it was found in Ref.~\cite{Guo:2020oqk} that the $D$-wave coupling accounts only for about half of the $D_1(2420)$ decay width, and a $D_1D^*\pi$ coupling in  $S$-wave is also required. The inclusion of an $S$-wave $D_1D^*\pi$ vertex could have a significant impact on the triangle diagram mechanism for the $\zc$ peak, since it provides a sizable different background contribution to the $J/\psi\pi$ invariant mass distribution. 
Both the $S$- and $D$-wave couplings are considered in Ref.~\cite{Du:2020vwb} in the tree-level diagrams, however there, the triangle  mechanism  is not included. Moreover, Gaussian regulators are employed in Refs.~\cite{Albaladejo:2015lob,Yang:2020nrt} to render the integrals in the Lippmann-Schwinger equation ultraviolet (UV) finite. While it is legitimate to use such a renormalization scheme in the small momentum regime, the interaction strength could be significantly enhanced (weakened) in the energy region far below (above) the relevant channel threshold. In order to bypass this issue and test the stability of the results obtained in Refs.~\cite{Albaladejo:2015lob,Yang:2020nrt}, we will use in this work dimensional regularization as only contact interactions are involved. 

The paper is organized as follows. In Sec.~\ref{sec:frame}, we briefly review the contact potentials, consistent with HQSS and SU(3) light-flavour symmetries, and  construct the hadron $T$-scattering matrix and production amplitudes taking both the point-like and triangle diagram production mechanisms into account. The parameters related to the $T$-matrix are determined from a best fit to the $\jp\pi$, $D\bar{D}^*$ and $D_s\bar{D}^*+D\bar{D}^*_s$ mass distributions in various scenarios and the resulting  poles are investigated in Sec.~\ref{sec:fit}. Section~\ref{sec:su3b} is devoted to studying possible SU(3) flavour violation effects. Finally,  the main conclusions of this work and a brief outlook are collected in Sec.~\ref{sec:con}. The explicit expression of the scalar three-point loop function is relegated to Appendix~\ref{sec:appen}.

\section{Formalism}\label{sec:frame}

In this section, we construct the coupled-channel amplitudes, considering both  $D\bar{D}^*$ ($D^*\bar{D}_s$ and $D\bar{D}_s^*$)  and  $\jp \pi$ ($\jp \bar K$) channels, which we use to describe the measured $D^0D^{*-}$ ($K^+$ recoil-mass ${\rm RM}(K^+)$) spectrum  in the $e^+e^-\to Y(4260)\to D\bar{D}^*\pi$~\cite{BESIII:2015pqw} ($e^+e^-\to K^+(D^{*0}\bar{D}_s+D^0\bar{D}_s^*)$~\cite{BESIII:2020qkh}) reaction and the $\jpi$ invariant mass distribution in the $e^+e^-\to\jp \pi\pi$ process~\cite{BESIII:2017bua}. To construct the $S$-wave contact potentials between the $D\bar{D}^*$, $D^*\bar{D}$, $D\bar{D}_s^*$, $D^{*}\bar{D}_s$, $\jpi$ and $\jp \bar K$ hadron pairs, we follow the approach of Ref.~\cite{Hidalgo-Duque:2012rqv}, which is  consistent with HQSS and SU(3) light-flavour symmetries.

\subsection{Contact interactions}

The two-particle $\left|D^{(*)}_{(s)} \bar D^{(*)}_{(s)}\right\rangle$ hadron states can be expanded in the HQSS basis  $\left|s_Q\otimes j_\ell;J\,I\,S\right\rangle$ \cite{Voloshin:2011qa}, with $s_Q$ and $j_\ell$ the total spin of the heavy-quark subsystem and the total angular momentum of the light degrees of freedom, respectively. For $S$-wave states, $j_\ell$ is just the total spin of the light-quark subsystem and the parity of the state is $P=+$. In addition, the total angular momentum, $J$, of the state is obtained by coupling  $s_Q\otimes j_\ell$, and $I$ and $S$ stand for the isospin\footnote{We use the isospin convention $\bar u
= \vert 1/2,-1/2\rangle$ and  $\bar d
= -\vert 1/2,+1/2\rangle$, which induces $D^0=\vert
1/2,-1/2\rangle$ and $D^+=-\vert
1/2,+1/2\rangle$.} and strangeness of the   light degrees of freedom. For simplicity, and when it cannot be misleading, we will omit the quantum numbers of the HQSS basis elements.  In what respects to the light degrees of freedom, the $D_{(s)}$ and $D_{(s)}^*$ form a $\left|\frac12\otimes\frac12\right\rangle$ spin multiplet. Thus in a given isospin-strangeness sector, the $S$-wave $\left|D^{(*)}_{(s)} \bar D^{(*)}_{(s)}\right\rangle$ can be decomposed as\footnote{The convention for charge conjugation used in this work is $\mathcal{C}D_{(s)}
\mathcal{C}^{-1} = \bar D_{(s)}  $ and $\mathcal{C}D^*_{(s)}\mathcal{C}^{-1}
=\bar D^*_{(s)}$. This convention is the same as in Ref.~\cite{Yang:2020nrt}, but differs by a minus sign in the transformation of the heavy-light vector meson field with respect to that used in Refs.~\cite{Hidalgo-Duque:2012rqv, Albaladejo:2015lob}.\label{footnote:C}}
\begin{subequations}
\begin{eqnarray}
\left(\begin{array}{c} \left|D_{(s) }\bar{D}_{(s)} \right\rangle\\  |D_{(s)}^* \bar{D}_{(s)}^*\rangle
\end{array}
\right)_{J=0} = \left(\begin{array}{cc}\frac12 & \frac{\sqrt{3}}{2} \\ \frac{\sqrt{3}}{2} & -\frac12 
\end{array}
\right)\left(\begin{array}{c}
|0\otimes 0\rangle \\ |1\otimes 1 \rangle
\end{array}\right)_{J=0},
\end{eqnarray}
\begin{eqnarray}
\left(\begin{array}{c} |D_{(s)}^* \bar{D}_{(s)}  \rangle \\ |D_{(s)} \bar{D}_{(s)}^* \rangle \\ |D_{(s)}^* \bar{D}_{(s)}^*  \rangle
\end{array}\right)_{J=1} = \left(\begin{array}{ccc}
-\frac12 & \frac12 & \frac{1}{\sqrt{2}}\\
\frac12 & -\frac12 & \frac{1}{\sqrt{2}} \\
\frac{1}{\sqrt{2}} & \frac{1}{\sqrt{2}} & 0 
\end{array}\right)\left(\begin{array}{c}
| 0\otimes 1\rangle \\ | 1\otimes 0\rangle \\ |1\otimes 1\rangle 
\end{array}\right)_{J=1},
\end{eqnarray}
\begin{eqnarray}
|D_{(s)}^* \bar{D}_{(s)}^* \rangle_{J=2} = |1\otimes 1\rangle_{J=2},
\end{eqnarray}
\end{subequations}
where the subindex $J$ denotes the total angular momentum.  HQSS implies that the strong interaction is independent of the conserved heavy quark spins in the $m_{Q}\to \infty$ limit. Therefore, we have
$\left\langle s_Q\otimes 1\left| \hat{\mathcal{H}}_I\right| s_Q\otimes 0\right\rangle =0$, and we introduce the  low energy constants (LECs) 
\begin{subequations}\begin{align}
 C_0 & \equiv \left\langle s_Q\otimes 0\left| \hat{\mathcal{H}}_I\right| s_Q\otimes 0\right\rangle,\\ 
C_1 & \equiv \left\langle s_Q\otimes 1\left| \hat{\mathcal{H}}_I\right| s_Q\otimes 1\right\rangle,
\end{align}\end{subequations}
for the interactions between the $D^{(*)}_{(s)}\bar{D}^{(*)}_{(s)}$ channels, where the LECs $C_{0,1}$ depend on the isospin and strangeness of the light-quark subsystem. We will come back to this point below since SU(3) flavor symmetry will reduce the number of independent LECs to only four~\cite{Hidalgo-Duque:2012rqv}.
Then the contact potentials have the form 
\begin{subequations}
\begin{align}
\label{eq:Vstates}
V_{J=0} & = \left( \begin{array}{cc}
\C_d+\frac12 \C_f & -\frac{\sqrt{3}}{2}\C_f \\
-\frac{\sqrt{3}}{2}\C_f & \C_d-\frac12\C_f
\end{array}\right) ,\\ \nonumber\\
V_{J=1} & = \left( \begin{array}{ccc}
\C_d+\frac12\C_f & \frac12\C_f & -\frac{1}{\sqrt{2}}\C_f \\
\frac12\C_f & \C_d+\frac12\C_f & \frac{1}{\sqrt{2}}\C_f \\
-\frac{1}{\sqrt{2}}\C_f & \frac{1}{\sqrt{2}}\C_f & \C_d
\end{array}\right) ,\\ \nonumber \\
V_{J=2} & = \C_d+\C_f~.
\end{align}
\end{subequations}
They coincide with the interactions derived from the effective Lagrangian in Ref.~\cite{Baru:2021ddn} by identifying $\C_d=(C_0+C_1)/2$ and $\C_f=(C_1-C_0)/2$, and with those obtained in  
Ref.~\cite{Hidalgo-Duque:2012rqv}, taking into account that in this latter case, there is a sign difference  between the $\bar{D}_{(s)}^*$ states used there and those employed in this work (see Footnote~\ref{footnote:C}). 

The charge conjugation of the $\left| D\bar D\right\rangle$, $\left| D_s\bar D_s\right\rangle$, $\left| D^*\bar D^*\right\rangle$ and  $\left| D_s^*\bar D_s^*\right\rangle$  $S$-wave  meson and anti-meson states  is $C=(-1)^J$. In addition, in  the $J=1$ sector, while the $C$-parity of the meson-antimeson pair $D^*\bar{D}^*$ can only be negative, both positive and negative $C$-parities can be achieved by the combinations
\bea 
\left| D\bar{D}^*\right\rangle_{J=1}^{C=\pm} = \frac{1}{\sqrt{2}}\left(\left|D\bar{D}^*\right\rangle \pm \left| D^*\bar{D}\right\rangle\right).
\eea
Thus, the potentials, $V_{J^{PC}}$, for $J^{PC}=1^{+-}$ and $J^{PC}=1^{++}$ read
\bea
V_{J^{PC}=1^{+-}} =\left(\begin{array}{cc} 
\C_d & \C_f \\
\C_f & \C_d
\end{array}\right), \quad V_{J^{PC}=1^{++}} = \C_d+\C_f,
\eea
with the basis $\left\{ \frac{1}{\sqrt{2}}\left(\left|D\bar{D}^*\right\rangle - \left| D^*\bar{D}\right\rangle\right), \left| D^*\bar{D}^*\right\rangle \right\}$ for $J^{PC}=1^{+-}$ and $\left\{ \frac{1}{\sqrt{2}}\left(\left|D\bar{D}^*\right\rangle + \left| D^*\bar{D}\right\rangle\right) \right\}$ for $J^{PC}=1^{++}$, respectively, which coincide with the results of Ref.~\cite{Hidalgo-Duque:2012rqv}. 

So far in the  discussion, we have not explicitly considered the  isospin-strangeness of the light quark subsystem. As mentioned above, the LECs $\C_{d,f}$ depend on these quantum numbers, but SU(3) symmetry provides relationships among them. From the light-quark flavor symmetry point of view, we have that three $\left|\bar D^{(*)0}\right\rangle,\left|\bar D^{(*)-} \right\rangle$ and $\left|\bar D^{(*)}_s \right\rangle$ anti-charmed heavy-light mesons form a SU(3) triplet ({\bf 3}) irreducible representation (irrep), which we denote as a column-vector  $\left|\bar{D}^{(*)A}\right\rangle, \, A=1,2$ and 3. Similarly, the  three $\left| D^{(*)0}\right\rangle,\left| D^{(*)+} \right\rangle$ and $\left| D^{(*)}_s \right\rangle$ charmed heavy-light mesons are the members of an SU(3) anti-triplet ({$\bf\bar 3$}) irrep, which we denote as a row-vector  $\left|D^{(*)}_A\right\rangle$. From the reduction ${\bf 3} \otimes {\bf \bar 3}= {\bf 1} \oplus {\bf 8}$, we conclude that assuming both HQSS and SU(3) flavor symmetries, there is only a total of four independent LECs, which correspond to SU(3) singlet and octet (or equivalently isoscalar and isovector) spin LECs ($C_0$ and $C_1$ or $\C_d$ and $\C_f$)~\cite{Hidalgo-Duque:2012rqv}. This is to say, for instance, $\C_f^{(8)}$,  $\C_d^{(8)}$,  $\C_f^{(1)}$  and  $\C_d^{(1)}$. In this notation, we can construct the singlet and octet representations as:
\begin{equation}
\left| \Ds\barDs,1\right\rangle  = \frac{1}{\sqrt{3}} \sum_{A}\left| \Ds_A\bar{D}^{(*)A}\right\rangle, \quad
 \left| \Ds\barDs,8; i\right\rangle  =  \frac{1}{\sqrt{2}} \sum_{A,B}(\lambda_i)^A_B \left| \Ds_A\bar{D}^{(*)B}\right\rangle ,
\end{equation}
where $\lambda_i$'s are the Gell-Mann matrices. It immediately follows
\bea
\left| \Ds_A\bar{D}^{(*)B}\right\rangle = \frac{1}{\sqrt{3}}\delta_A^B \left| \Ds\barDs,1\right\rangle +
\frac{1}{\sqrt{2}}\sum_{i=1}^8 (\lambda_i)_A^B \left| \Ds\barDs,8; i\right\rangle.
\eea
 As the SU(3) symmetry implies that the $\Ds\barDs$ interaction only distinguishes the singlet and octet representations, we introduce 
\bea
\left\langle \Ds\barDs,1\left|\hat{\mathcal{H}}_I\right|  \Ds\barDs,1\right\rangle = C^{(1)},\quad 
\left\langle \Ds\barDs,8;i\left|\hat{\mathcal{H}}_I\right|  \Ds\barDs,8;j\right\rangle = C^{(8)}\delta_{ij},
\eea
where $C^{(1)}$ and $C^{(8)}$ rely on the heavy-quark structure of the system. Then it is straightforward to obtain (without summing over either $A$ or $B$ in the following expressions and for $A\neq B$)
\bea
\left\langle \Ds_A\bar{D}^{(*)A} \left| \hat{\mathcal{H}}_I\right|\Ds_A\bar{D}^{(*)A}\right\rangle & = &  \frac13 (C^{(1)}+2C^{(8)}), \nonumber\\
\left\langle \Ds_A\bar{D}^{(*)A} \left| \hat{\mathcal{H}}_I\right|\Ds_B\bar{D}^{(*)B}\right\rangle & = &   \frac13 (C^{(1)}-C^{(8)}),  \nonumber\\ 
\left\langle \Ds_A\bar{D}^{(*)B} \left| \hat{\mathcal{H}}_I\right|\Ds_A\bar{D}^{(*)B}\right\rangle & = &  C^{(8)}, 
\eea
The octet representation contains an isospin triplet ($S=0,I=1$), one isospin singlet ($S=0,I=0$) and two isospin doublets ($S=\pm 1,I=\frac12$). It follows that in the basis $\left|\Ds\barDs,8;S\,I\,M_I\right\rangle$,
\bea
\left\langle \Ds\barDs,8;0\,1\,M_I\left|\hat{\mathcal{H}}_I\right|  \Ds\barDs,8;0\,1\,M_I\right\rangle 
&=&\left\langle \Ds\barDs,8;\pm1\,\frac12\, M_I\left|\hat{\mathcal{H}}_I\right|  \Ds\barDs,8;\pm 1\,\frac12\, M_I\right\rangle\nonumber\\ 
&=& C^{(8)},
\eea
and hence we find that the interactions 
in the channels of the $\zc$ and of its strange partner [$\zcs$] are identical in the SU(3) limit~\cite{Yang:2020nrt}. 
As a result, one finds in the heavy quark limit that for the $J^P=1^+$ sector
\bea\label{eq:Vdd}
V^{I=1}_{J^{PC}=1^{+-}}[D\bar{D}^*] = V^{I=\frac12}_{J^P=1^+}[D\bar{D}^*_s]= V^{I=\frac12}_{J^P=1^+}[D_s\bar{D}^*]  = \left(\begin{array}{cc} 
\C_d^{(8)} & \C_f^{(8)} \\
\C_f^{(8)} & \C_d^{(8)} \label{eq:VHQSS}
\end{array}\right),
\eea
with the bases in the order $\left\{ \frac{1}{\sqrt{2}}\left(\left|D\bar{D}^*\right\rangle - \left| D^*\bar{D}\right\rangle\right), \left| D^*\bar{D}^*\right\rangle \right\}$, $ \left\{ \frac{1}{\sqrt{2}}\left(\left|D\bar{D}^*_s\right\rangle - \left| D^*\bar{D}_s\right\rangle\right), \left| D^*\bar{D}^*_s\right\rangle \right\}$  and $\left\{ \frac{1}{\sqrt{2}}\left(\left|D_s\bar{D}^*\right\rangle - \left| D^*_s\bar{D}\right\rangle\right), \left| D^*_s\bar{D}^*\right\rangle \right\}$, respectively. One can make contact with the results of Ref.~\cite{Hidalgo-Duque:2012rqv} using  $\C_d^{(8)}= (C_{1a}-C_{1b})$ and $\C_f^{(8)}=2C_{1b}$ with $C_{1a}$ and $C_{1b}$ the LECs defined therein. Note that, assuming the existence of the  $\zc$ and of the $\zcs$, then the structure of the potentials of Eq.~\eqref{eq:VHQSS} supports the existence of their $D^*\bar D^*$ and  $D^*\bar{D}^*_s$ HQSS partners, where the first one could be identified with  the $Z_c(4025)$, and the second one was first predicted in Ref.~\cite{Yang:2020nrt}.

The inelastic transitions between the $D_{(s)}\bar{D}_{(s)}^*$ and $\jp \pi$ ($\jp  K$, $\jp \bar K$) are also required to describe the $\jp\pi$ mass distribution, as well as to account for the contribution from $\jp\pi$ and $\jp \bar K$ to the widths of the $Z_c$ and $Z_{cs}$, respectively.\footnote{In addition to the $D_{(s)}\bar{D}_{(s)}^*$ and $\jp \pi (\bar K)$ channels, the $\eta_c\rho (\bar K^*)$ threshold is located in the energy range of interest. However, regarding the $Z_{c(s)}$ structures and the $D_{(s)}\bar{D}_{(s)}^*$ and $\jp \pi$ invariant mass distributions,  neglecting  the $\eta_c\rho(K^*)$ channel is a sounded approximation. On one hand, the $\eta_c\rho (\bar K^*)$ interaction is Okubo-Zweig-Iizuka (OZI) suppressed and no significant structure is observed near its threshold. On the other hand, the $\eta_c\rho (\bar K^*)$ threshold is relatively far from the $Z_{c(s)}$ peak and the $J/\psi\pi (\bar K)$ threshold, which makes it possible to absorb, in  an effective way,  its contribution in the inelastic $D_{(s)}\bar{D}_{(s)}^*\to J/\psi\pi (\bar K)$ transition contact term. In principle, the channel $\eta_c\rho (\bar K^*)$ could also be included explicitly, once the data of interest for this channel were statistically significant.
However, currently the signal of the $Z_c(3900)$ in the $\rho\eta_c$ channel~\cite{BESIII:2019rek} is much less significant/important than that in the $J/\psi\pi$ channel. Nevertheless, it is worth exploring whether the HALQCD observation that the $\eta_c\rho$ channel has a strong coupling with the $D\bar D^*$ channel~\cite{HALQCD:2016ofq} is consistent with this experimental fact.} The $S$-wave $\jp \pi$ ($\jp K$, $\jp \bar K$ ) system can also be expressed in the heavy-light basis $\left| \jp \pi(\bar K, K)\right\rangle_{J=1}=\left|s_Q= 1\otimes j_\ell=0\right\rangle^\prime $, with the isospin and strangeness of the state determined by the Golsdtone boson (pion, antikaon or kaon) of the state. Following the same strategy discussed above, it is straightforward to find that $\jp \pi$ ($\jp \bar K$) only couples to   $\frac{1}{\sqrt{2}}\left(D\bar{D}^*_{(s)}-D^*\bar D_{(s)}\right)$ and $D^*\bar D^*_{(s)}$ in $J=1$  and thus using the same elements of the basis as in Eq.~\eqref{eq:VHQSS}, we find
\bea
\label{eq:Vjp} 
V^{I=1}_{J^{PC}=1^{+-}}[D\bar{D}^*\to \jp \pi] = V^{I=\frac12}_{J^P=1^+} [D\bar{D}^*_s\to \jp  \bar K]= V^{I=\frac12}_{J^P=1^+}[D_s\bar{D}^*
\to \jp K] = \left(\begin{array}{c}
\mathcal{V}\\ {-} \mathcal{V}
\end{array}\right),~~
\eea
where we have introduced the LEC $\mathcal{V} = -\left\langle 1\otimes 0\left|\hat{\mathcal{H}}_I\right| 1\otimes 0\right\rangle^\prime/{\sqrt{2}} $ and assumed SU(3) flavor symmetry.  The direct transitions $\jp \pi\to \jp \pi$, $\jp \bar K\to \jp \bar K$ and  $\jp K\to \jp K$ are  neglected due to the OZI suppression. The contact interaction potentials can also be constructed from an effective Lagrangian; see, {\it e.g.}, Refs.~\cite{Mehen:2011yh,Baru:2021ddn}.

In this work, we will focus on the $\zc$ and $\zcs$ resonances, which are close to the  $D\bar{D}^*$ and $D\bar{D}^*_s$/$D^*\bar{D}_s$ thresholds, respectively. To reduce the number of free parameters, we neglect the $D^*_{(s)}\bar{D}_{(s)}^*$ higher coupled-channel, and we only  include explicitly  $\jp \pi$ and $\left(D\bar{D}^*-D^* \bar D\right)/ \sqrt{2}$, or  $\jp \bar K$ and $\left(D\bar{D}^*_s-D^*\bar{D}_s\right)/\sqrt{2}$, labelled as channels 1 and 2, respectively. As commented above, the first channel acts as a support or decay channel. Then the couple-channel contact interactions in the $\zc$ and $\zcs$ sectors read 
\bea
\label{eq:Vs}
V^{I=1}_{J^{PC}=1^{+-}} = V^{I=\frac12,s}_{J^P=1^+} =\left(\begin{array}{cc}
0& V^{(s)}_{12}\\ V^{(s)}_{12} & V^{(s)}_{22} 
\end{array}\right) = \left(\begin{array}{cc}
0 & \mathcal{V} \\ \mathcal{V} & \C_d^{(8)}
\end{array}\right),
\eea
where the superscript $s$ stands for the strange sector. In what follows, we parameterize the matrix elements of the potential as\footnote{Here the $\sqrt{2m_\pi}$ factor is introduced only for dimensional arguments, and does not imply that we treat  non-relativistically the pion.} 
\bea\label{eq:Cs}
V_{12}^{(s)} = \sqrt{2m_{D_{(s)}}2m_{D^*}2m_{\jp}2m_\pi} \tilde{V}_{12}^{(s)},
\qquad V_{22}^{(s)} = 2m_{D_{(s)}}2m_{D^*}\tilde{V}_{22}^{(s)}.
\eea 
HQSS and SU(3) light flavor symmetries imply that  $\tilde{V}^s_{12}=\tilde{V}_{12}$ and $\tilde{V}_{22}^s=\tilde{V}_{22}$.\footnote{Note that the use of $m_D$ or $m_{D_s}$ in Eq.~\eqref{eq:Cs}, from the normalization pre-factors of the heavy-fields, introduces a small SU(3) breaking correction.} The coupled-channel $T$-matrix can be obtained by 
\bea\label{eq:Tmatrix}
T^{(s)} = \frac{1}{1-V^{(s)}\cdot G^{(s)}}V^{(s)},
\eea
where $G^{(s)}$ is the loop-function diagonal matrix $G(s)=\text{diag} \left\{G_1(s),G_2(s)\right\}$, 
\bea\label{eq:Gintegral}
G_i(s) = i\int\frac{d^4q}{(2\pi)^4}\frac{1}{(q^2-m_{i,1}^2+i\epsilon)\left[(p-q)^2-m_{i,2}^2+i\epsilon\right]}, \quad s= p^2,
\eea
with $m_{i,1}$ and $m_{i,2}$ the masses of the two mesons in the $i$th channel. Note that the use of physical masses in the above loop functions breaks SU(3) symmetry. The loop functions are logarithmically divergent and need to be regularized. In Refs.~\cite{Albaladejo:2015lob,Yang:2020nrt}, Gaussian form factors are introduced into the potentials to render the $G_i(s)$ function UV well defined, which however distorts the interaction strength for energies far from thresholds. In this work, we evaluate the loop function $G_i(s)$ with a once-subtracted dispersion relation and its explicit expression reads for  $s\ge(m_{i,1}+m_{i,2})^2$~\cite{Oller:1998zr}
\bea
{\rm Re}\,G_i(s) & = & \dfrac{1}{16\pi^2}\Bigg[  a_i(\mu) + \log\frac{m_{i,1}^2}{\mu^2} + \frac{s-m_{i,1}^2+m_{i,2}^2}{2s}\log \frac{m_{i,2}^2}{m_{i,1}^2}\nonumber \\
&& + \frac{\sigma_i(s)}{2s}\log\frac{s+\sigma_i(s)-m_{i,1}^2-m_{i,2}^2}{s-\sigma_i(s)-m_{i,1}^2-m_{i,2}^2} \Bigg], \nonumber\\
{\rm Im}\,G_i(s) & = & -\frac{\sigma_i(s)}{16\pi s},
\label{eq:G}
\eea
with $\sigma_i(s) =  [s-(m_{i,1}+m_{i,2})^2]^\frac12[s-(m_{i,1}-m_{i,2})^2]^\frac12$ and $\mu$ the renormalization scale. The $G_i(s)$ function should be $\mu$-independent, and a change of $\mu$ should be compensated by that of $a_i(\mu)$. The value of $a_i(\mu)$ could  be estimated by matching the above expression for $G_i(s)$ to the loop function regularized by a hard cutoff $q_{\rm max}$ of the order of 1-1.5 GeV (see for instance, Eqs.~(51) and (52) of Ref.~\cite{Garcia-Recio:2010enl}). Taking into account that we have to evaluate the function $G_i(s)$ not only for real $s\ge(m_{i,1}+m_{i,2})^2$, but also below threshold and in the second Riemann sheet (RS) of the complex-$s$ plane
as well, to look for the position of resonances, we refer to Appendix~A of Ref.~\cite{Nieves:2001wt} for its analytical continuation.

\subsection{Production amplitudes}\label{subsec:amplitudes}

The $T$-matrix introduced in Eq.~\eqref{eq:Tmatrix} accounts for the final-state re-scattering, and one additionally needs to construct a model for the production amplitudes for the $e^+e^-\to \jp \pi\pi$, $e^+e^-\to  D\bar{D}^*\pi$, and $e^+e^-\to K^+D^{*0}\bar{D}_s/K^+D^0 \bar{D}_s^*$ reactions. 
In Refs.~\cite{Wang:2013cya,Wang:2013hga,Albaladejo:2015lob,Gong:2016jzb,Pilloni:2016obd,Guo:2020oqk}, it was demonstrated that the $D_1(2420)\bar{D}D^*$ triangle diagram is important for the understanding of the $\zc$ . It can produce a resonance-like structure, which enhances the production of near-threshold resonances~\cite{Guo:2019twa}. In addition to the triangle-diagram production mechanism, we will also include the direct point-like production in this work. The Feynman diagrams for the $e^+e^-\to \jp \pi\pi$ and $D\bar{D}^*\pi$ are shown in Fig.~\ref{fig:feyn_zc}, where the reactions proceed through the formation of the resonance $Y(4230)/Y(4260)$. The value of the coupling $YD_1D$ is irrelevant for the description of the line shapes as it can be effectively absorbed into the overall normalization factors. 
Regarding the $D_1D^*\pi$ vertex, in Refs.~\cite{Wang:2013cya,Albaladejo:2015lob} a $D$-wave coupling is used supported by the small width of $D_1(2420)$, $(31.3\pm1.9)$~MeV~\cite{ParticleDataGroup:2020ssz}, which suggests that it is approximately a charmed meson with $j_\ell^P=\frac32^+$, with $j_\ell$ the total angular momentum of its light degrees of freedom, and thus decays into the $D^{(*)}\pi$ mainly in $D$-wave. The $D$-wave $D_1D^*\pi$ coupling is described by the Lagrangian 
\bea\label{eq:lag_D}
\lag_D = \frac{h_D}{2F_\pi}\text{Tr}\big[ T_b^i\sigma^jH_a^\dag\big]\partial^i\partial^j\phi_{ba},
\eea
which respects HQSS~\cite{Casalbuoni:1996pg,Guo:2020oqk}, where $F_\pi=92.1 $ MeV and  $\sigma^i$ are the pion decay constant and the spin Pauli matrices, respectively. On the other hand, the super-fields $H_a$ and 	$T_a^{i}$,
\bea H_a = D_a^{*i}\sigma^i+D_a,\qquad	T_a^{i} = D_{2a}^{ij}\sigma^j+\sqrt{\frac{2}{3}}D_{1a}^i+i\sqrt{\frac16}\epsilon_{ijk}D_{1a}^j\sigma^k,
\eea
represent the $j_\ell = \frac12^-$ and $\frac32^+$ spin multiplets, respectively. In addition,  Tr[$\cdot$] denotes the trace in the spinor space, and $\phi_{ba}$ collects the pion fields with $a$, $b$ the light flavor indices 
\bea
\phi = \left( \begin{array}{cc}
\pi^0/\sqrt{2} & \pi^+ \\ \pi^- & -\pi^0/\sqrt{2}
\end{array}\right).
\eea
The  $D$-wave coupling $h_D$ is determined from the central value of the $D_2$ width to be $|h_D|=1.17~\text{GeV}^{-1}$~\cite{Guo:2020oqk}, which in turn leads to 15.2 MeV for the $D_1(2420)$ width, that is, only about half of the total. By requiring that the decay of the $D_1$ is saturated by the $D^*\pi$ (and sequential $D\pi\pi$) mode, an $S$-wave $D_1D^*\pi$ coupling is required to account for the rest of the $D_1$ width~\cite{Guo:2020oqk},
\bea
\lag_S = i \frac{h_S}{\sqrt{6}F_\pi}D_{1b}^i D_a^{*i\dag}\partial^0\phi_{ba},
\eea
with $|h_S|=0.57$, where the  pion-energy factor is introduced because the pions are pseudo-Goldstone bosons of the spontaneous breaking of chiral symmetry.

\begin{figure*}[tb]
 \centering
  \includegraphics[width=1.0\textwidth]{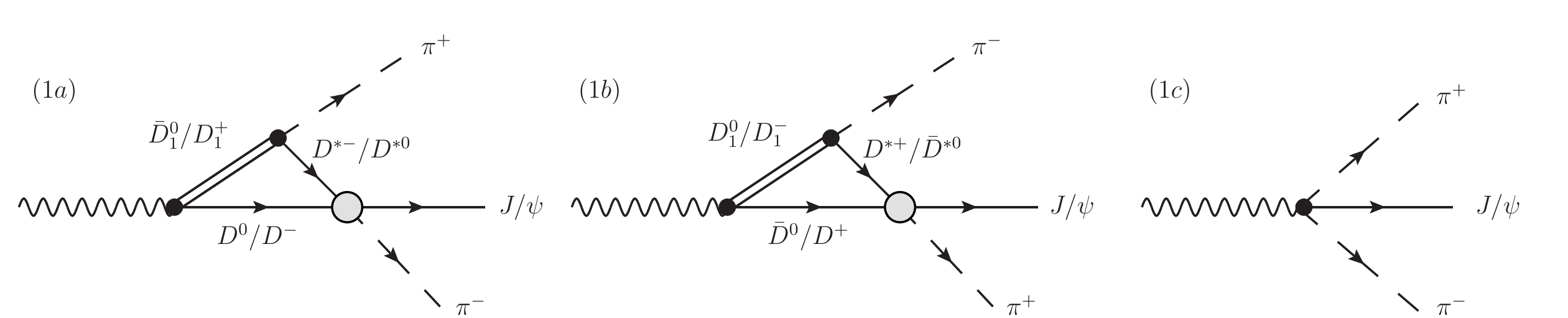}
  \includegraphics[width=1.0\textwidth]{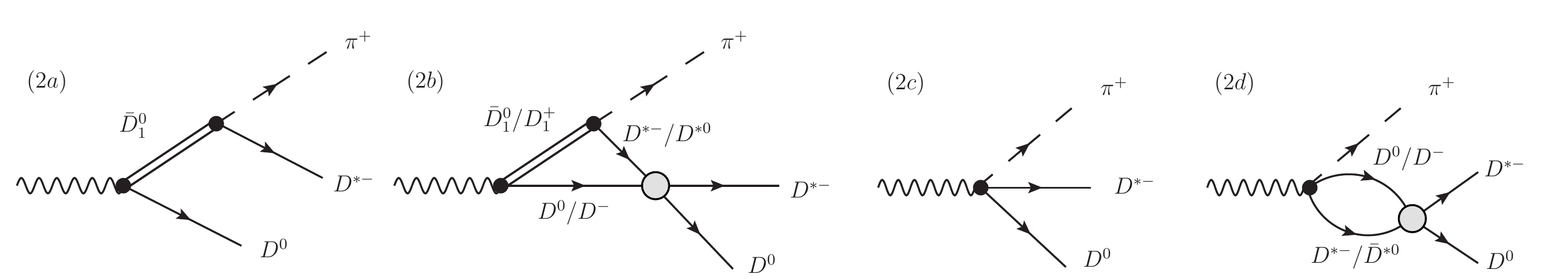}
  \caption{Mechanisms for the $Y(4230/4260)\to \jp\pi\pi$ and  $Y(4230/4260)\to \pi\bar{D}^*D$ reactions. In the first case, the panels (1a)--(1b) and (1c) stand for the $D_1\bar{D}D^*$ triangle diagram and the direct point-like  productions of $\jp\pi\pi$, respectively. The production of the $\pi\bar{D}^*D$ through the intermediate  $D_1$ state and  the point-like production followed by the final-state re-scatterings are shown in panels (2a)--(2d).}
  \label{fig:feyn_zc}
\end{figure*}

We denote by $\A_1(s,t)$ and $\A_2(s,t)$ the amplitudes of the  $Y\to \jp\pi^+ \pi^-$ and $Y\to \pi^+ D^{*-}D^0$ decays, respectively. Here $s$ ($t$) stands for the square of the invariant masses  of the  $\jp\pi^-$ or $D^{*-}D^0$ ($\jp \pi^+$ or $D^{*-}\pi^+$) pairs, respectively, for each of the two reactions. Then, up to some irrelevant constants, the amplitude $\A_1(s,t)$ reads 
\bea\label{eq:ampA1}
\A_1(s,t) &=& \epsilon_Y^i\epsilon_{\jp}^{*j}\left \{\left( 3q^i_+q^j_+ - \delta^{ij}q^2_+ \right) \left[I(s)T_{12}(s)+\frac{\alpha}{\sqrt{2}}\right] + \frac{h_S}{h_D}E_{\pi^+} I(s)T_{12}(s) \right\} \nonumber\\
& & + (s\leftrightarrow t, q_+\leftrightarrow q_-, E_{\pi^+}\leftrightarrow E_{\pi^-}),
\eea
where $q_{\pm}$ and $E_{\pi^{\pm}}$ are the three-momentum and the energy of the $\pi^{\pm}$, respectively, and the $\alpha$ term accounts for the $D$-wave contribution of the point-like production (panel (1c) in Fig.~\ref{fig:feyn_zc}). The $S$-wave $\pi^+\pi^-$ re-scattering is not explicitly included  in the amplitude and its contribution will be modeled by a symmetric smooth background. On the other hand, $T_{12}(s)$ represents the scattering amplitude of $D\bar{D}^*\to \jp\pi^-$ obtained in Eq.~\eqref{eq:Tmatrix} and $I(s)$ is the scalar three-point $(D_1\bar{D}D^*)$ loop function 
\bea\label{eq:3-point}
I(s) = i\int\frac{d^4q}{(2\pi)^4} \frac{1}{\left(q^2-m_{D_1}^2\right)\left[(P-q)^2-m_D^2\right]\left[(q-k)^2-m_{D^*}^2\right]}, 
\eea
where $P^2=M^2, k^2=m_\pi^2, s=(P-k)^2$ with $M$ the total c.m. energy of $e^+e^-$ system. The expression for the triangular  loop function of Eq.~\eqref{eq:3-point} is explicitly given in Appendix~\ref{sec:appen}. After the appropriate sum and average over the polarizations, one obtains\footnote{There is a typo in Ref.~\cite{Albaladejo:2015lob} for the corresponding equation. The factor $1/4$ of the third term on the right side of Eq.~(6) in that reference should be $1/2$ (c.f. the third term in Eq.~\eqref{eq:absA1} here). We have checked that it has only a marginal numerical impact.}
\bea
\overline{\left| \A_1(s,t) \right|^2} &=& |\tau(s)|^2 q_\pi^4(s) + |\tau(t)|^2 q_\pi^4(t) + 
\frac{3\cos^2\theta-1}{2}\big[\tau(s)\tau(t)^*+\tau(s)^*\tau(t)\big]q_\pi^2(s)q_\pi^2(t) \nonumber\\
&& + \frac{1}{2} \Big\{ |\tau^\prime(s)|^2 E_\pi^2(s) + |\tau^\prime(t)|^2 E_\pi^2(t) + \big[\tau^\prime(s)^*\tau^\prime(t) + \tau^\prime(s)\tau^\prime(t)^* \big] E_\pi(s)E_\pi(t) \Big\},
\label{eq:absA1}
\eea
where $q_\pi^2(s)= \lambda(M^2,s,m_\pi^2)/(4M^2)$ with $\lambda(x,y,z)=x^2+y^2+z^2-2xy-2yz-2xz$ the K\"all\'en function, $E_\pi(s)=(M^2+m_\pi^2-s)/(2M)$, and $\theta$ is the relative angle between the two pions in the $Y(4230)/Y(4260)$ rest frame. In addition, 
\begin{equation}
\tau(s) = \sqrt{2}I(s)T_{12}(s)+\alpha, \qquad \tau^\prime(s) = \frac{h_S}{h_D}\sqrt{2}I(s)T_{12}(s).
\end{equation}
Likewise, one finds
\bea\label{eq:ampA2}
\overline{| \A_2(s,t)|^2} & = & \left| \frac{1}{t-m_{D_1}^2} + I(s)T_{22}(s)\right|^2 q_\pi^4(s) \nonumber\\
& &  +  \frac12 \left| E_\pi(s)\frac{h_S}{h_D} \left[\frac{1}{t-m_{D_1}^2} + I(s)T_{22}(s) \right] + \beta\Big[1+G_2(s)T_{22}(s)\Big]\right |^2,
\eea
where $\beta$ accounts for the $\pi^+{D}^{*-}D^0$ point-like production, {\it i.e.} panel (2$c$) of Fig.~\ref{fig:feyn_zc}. In Ref.~\cite{Albaladejo:2015lob}, only the $D$-wave $D_1\to D^*\pi$ transition was considered. To assess the impact of the $S$-wave vertex, we will consider  in this work two cases: only $D$-wave, {\it i.e.} $h_S=0$, as in Ref.~\cite{Albaladejo:2015lob}, and both $S$- and $D$-wave couplings.

The data on $e^+e^-\to K^+(D^0 \bar D_s^*+D^{*0}\bar D_s)$ are measured within the $e^+e^-$ c.m. energy region $M\in$~[4.268, 4.698]~GeV \cite{BESIII:2020qkh}. In this energy region, especially for $M=4.681$ GeV, it is found that the {$D_{s2}^*(2573)\bar{D}^*_sD^0$} triangle diagrams, see Fig.~\ref{fig:feyn_zcs}, can facilitate the production of the $\zcs$ resonance. The $D_{s2}^*DK$ vertex can be described by the SU(3) extension of the Lagrangian of Eq.~\eqref{eq:lag_D}. Since only the $D$-wave coupling is involved for $D_{s2}^*\to D^{(*)}K$, its  strength is irrelevant as it can be reabsorbed into the normalization factor for line shapes. 
For the $e^+e^-$ energy values  measured by BESIII, it is natural to assume that the reaction $e^+e^-\to K^+(D^0\bar{D}_s^*+D^{*0}\bar{D}_s )$ 
 proceeds through the $\psi(4660)$ resonance,  $e^+e^-\to\psi(4660)\to   K^+(D^0\bar{D}_s^*+D^{*0}\bar{D}_s )$. We denote by $\A_{s,1}(s,t)$ [$\A_{s,2}(s,t)$] the amplitude for $\psi(4660)\to K^+D^0\bar{D}_s^*$ [$\psi(4660)\to K^+D^{*0}\bar{D}_s$], with $s$ and $t$ the $D^0\bar{D}_s^*$ and $D^0K^+$ [$D^{*0}\bar{D}_s$ and $D^{*0}K^+$] invariant masses squared, respectively. We  obtain, up to an irrelevant common constant~\cite{Yang:2020nrt},\footnote{The contributions of the production through $D_{s2}^*$ and the point-like one in Ref.~\cite{Yang:2020nrt} should be added incoherently, instead of coherently as done in Ref.~\cite{Yang:2020nrt}, because their accompanied $K^+$ are in $D$- and $S$-waves, respectively. The impact on the numerical results of Ref.~\cite{Yang:2020nrt} is marginal due to the very small contribution from the interference terms.}
\bea\label{eq:ampAs}
\overline{\left| \A_{s,1}(s,t)\right|^2} &=& \left| \frac{1}{t-m_{D_{s2}^*}^2+im_{D_{s2}^*}\Gamma_{D_{s2}^*}} + \frac12 I_s(s)T_{22}^s(s) \right|^2 q_K^4(s)\nonumber \\
&&+ r^2\left|  1 + \frac12 \left[G_{D\bar{D}_s^*}(s) + G_{D^*\bar{D}_s}(s)\right]T_{22}^s(s) \right|^2,\nonumber\\
\overline{\left| \A_{s,2}(s,t)\right|^2} &=& \left|\frac12 I_s(s)T_{22}^s(s)\right|^2 q_K^4(s)+ r^2 \left|1+ \frac12 \left[G_{D\bar{D}_s^*} (s)+ G_{D^*\bar{D}_s}(s)\right]T_{22}^s(s) \right|^2,
\eea
where $q_K^2(s)= \lambda(M^2,s,m_K^2)/(4M^2)$, $G_{P_1P_2}$ is the loop function of Eq.~\eqref{eq:G} evaluated with $P_1$ and $P_2$ the 
particles running in the two-body loop, and $I_s(s)$ represents the $D_{s2}^*\bar{D}_s^*D$ scalar triangle integral. Here we have used the relation 
\begin{equation}
    T_{D^0\bar{D}_s^*\to D^0\bar{D}_s^*}(s) = -T_{D^0\bar{D}_s^*\to D^{*0}\bar{D}_s}(s)= \frac{T^s_{22}(s)}{2} ,
\end{equation}
and $r$ represents the point-like production of $K^+D^0\bar{D}_s^*$/$K^+ D^{*0}\bar{D}_s$ and is a parameter accounting for the relative weight between diagrams (3a, 3b) and 
(3c, 3d) in Fig.~\ref{fig:feyn_zcs}.

\begin{figure*}[tb]
 \centering
  \includegraphics[width=1.0\textwidth]{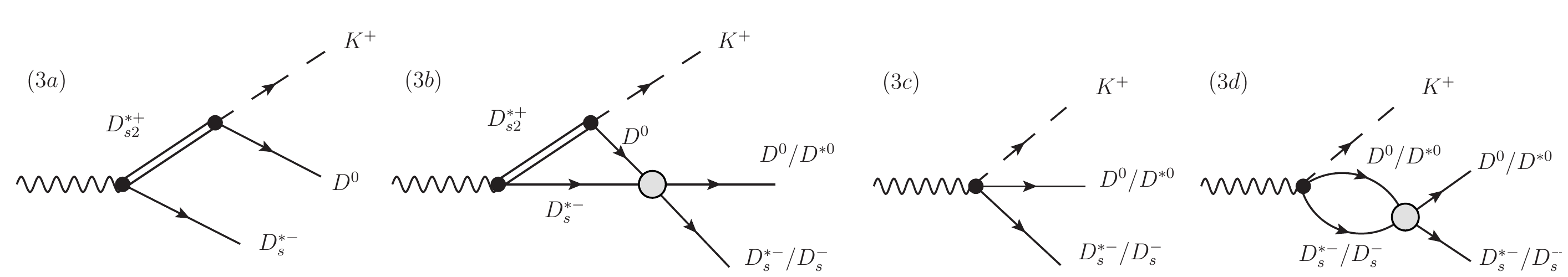}
  \caption{Diagrams for the $e^+e^-\to K^+(D^0\bar{D}_s^*+D^{*0}\bar{D}_s )$.  }
  \label{fig:feyn_zcs}
\end{figure*}

For the  $Y\to \jp\pi^+ \pi^-$ and $Y\to \pi^+ D^{*-}D^0$ reactions, the $\jp\pi^-$ and $D^0 D^{*-}$ spectra are obtained from the amplitudes as
\bea\label{eq:linezc}
\frac{d\Gamma_i}{d\sqrt{s}} = \mathcal{N}_i\sqrt{s}\int_{t_i^-}^{t_i^+} dt\, \overline{\left|\A_i(s,t)\right|^2} + \mathcal{B}_i(s),\qquad i=1,2
\eea
where $\mathcal{N}_i$ is an unknown normalization factor,  $t_i^{\pm}(s)$ are the limits of the Mandelstam variable $t$ for the decay mode $i$ (see, {\it e.g.}, the {\it Dalitz plot} section of the {\it Kinematics} review in Ref.~\cite{ParticleDataGroup:2020ssz}), and $\mathcal{B}_i(s)$ is an incoherent background mimicking possible contributions from crossed channels, misidentified events, and higher waves other than the $S$-wave. For the $\jp\pi^-$ and $D^0 D^{*-}$ spectra, the background is parameterized as~\cite{Albaladejo:2015lob,BESIII:2013ris} 
\bea
\mathcal{B}_i(s) = B_i\left[ (\sqrt{s}-m_{i,-})(m_+-\sqrt{s})\right]^{d_i}, \label{eq:bkg}
\eea
with $m_{1,-}=m_{\jp}+m_\pi$, $m_{2,-}=m_{D^*}+m_D$, and $m_+=M-m_\pi$. In this work, we will fix $d_i=1$ to reduce the number of free parameters. We have checked that releasing $d_i$ as a free parameter barely improves the fit quality. 

For the {$e^+e^-\to K^+(D^0 \bar D_s^*+D^{*0}\bar D_s)$}, we will be only interested in the energy region close to the {$D^*\bar D_s$ and $D \bar D_s^*$} thresholds and we will only fit to the low-energy tail. Hence, we will simply use the background {($\mathcal{B}$)}  employed in the experimental analysis of Ref.~\cite{BESIII:2020qkh}. The line shape for the $K^+$ recoil mass distribution at the different total $e^+e^-$ energy values $M$ can be computed as~\cite{Yang:2020nrt}
\begin{equation}
\frac{dN_s}{d\sqrt{s}} = \mathcal{N}_s\sqrt{s}\lag_\text{int}f_\text{corr}\bar{\epsilon}\left|\frac{1}{M^2-m_\psi^2+im_\psi \Gamma_\psi}\right|^2\left( \int_{t_{s,1}^-}^{t_{s,1}^+}dt\, \overline{\left| \A_{s,1}(s,t) \right|^2} 
+ \int_{t_{s,2}^-}^{t_{s,2}^+}dt\, \overline{\left| \A_{s,2}(s,t) \right|^2}\right) {+\mathcal{B}},
\label{eq:linezcs}
\end{equation}
with $t_{s,i}^{\pm}(s)$ the limits of the Mandelstam variable $t$ for the decay mode $i$. The integrated luminosity $\lag_\text{int}$, the detection efficiency $\bar{\epsilon}$ and the correction factor $f_\text{corr}$ can be found in Ref.~\cite{BESIII:2020qkh}, and $m_\psi=4630$~MeV and $\Gamma_\psi=62$~MeV are employed for the mass and width of the $\psi(4660)$ resonance, respectively~\cite{ParticleDataGroup:2020ssz}. {Finally, the factor $  \mathcal{N}_s$ is associated with the
$e^+e^-$ annihilation vertex, and here we take the same value for the
$e^+e^-$ energy range from 4.628 to 4.698~GeV}.

\section{Description of the experimental spectra: the  SU(3) flavour symmetric case}\label{sec:fit}

In this section, we will determine the hidden-charm two-meson scattering  $T$-matrix of Eq.~\eqref{eq:Tmatrix}, assuming SU(3) flavour symmetry, from a combined fit to the $\jp\pi^-$, $D^0 D^{*-}$, and the $K^+$ recoil-mass
[${\rm RM}(K^+)$] distributions of the $e^+e^-\to \jp\pi^+\pi^-$ at c.m. energies of 4.23 and 4.26 GeV~\cite{BESIII:2017bua}, $e^+e^-\to \pi^+D^0D^{*-}$ at 4.26 GeV~\cite{BESIII:2015pqw}, and $e^+e^-\to K^+(D^{*0} D_s^-+D^0 D_s^{*-})$ at 4.628, 4.641, 4.661, 4.681 and 4.698 GeV~\cite{BESIII:2020qkh}. 
The leading-order (LO)  approximation for the contact potentials of Eq.~\eqref{eq:Vs}  amounts to approximating them simply by constants. However, such interaction kernels cannot generate  resonances above the higher threshold, even when coupled channel effects are considered.  To avoid this problem and to take into account the possibility of the $Z_{c(s)}$ poles being located above the corresponding open-charm thresholds, we introduce an energy-dependent term in $\tilde{V}^{(s)}_{22}$, and fix $\tilde{V}^{(s)}_{12}$ to a constant, since the $\jp\pi$ ($\jp \bar K$) threshold is located far from  the $Z_c$ ($Z_{cs}$),
\bea
\label{eq:V22}
\tilde{V}^{(s)}_{22} = C_Z + \frac{b}{2(m_{D_{(s)}}+m_{D^*})}\left[ s-(m_{D_{(s)}}+m_{D^*})^2\right], \qquad \tilde{V}^{(s)}_{12}= C_{12},
\eea
where $C_Z$, $b$, and $C_{12}$ are constants. The LO constant potential case corresponds to $b=0$, and  we will consider both scenarios below, {\it i.e.}, $b=0$ and $b\neq 0$. Before proceeding to describe the data, we need to discuss the subtraction constants $a_i(\mu)$ in the $G_i(s)$ functions defined in Eq.~\eqref{eq:G}, which can be determined by matching the loop function evaluated at threshold with a hard cutoff or a Gaussian form factor with the cutoff $\Lambda$  around 1~GeV~\cite{Guo:2006fu,Garcia-Recio:2010enl}.  We have taken $a_i(\mu)=a_i^s(\mu)$, consistent with SU(3) flavor symmetry. Nevertheless, it is easy to check that variations of $a_1(\mu)$ can be fully absorbed into changes of the  $C_Z$, $b$, and $C_{12}$ LECs. However, this is not the case for the subtraction constant $a_2(\mu)$.\footnote{The reason is that we have set the contact term for the diagonal $J/\psi\pi\,(J/\psi K)\to J/\psi\pi\,(J/\psi K)$ element of the potential matrix to zero (see Eq.~\eqref{eq:Vs}). It was shown in Refs.~\cite{Cohen:2004kf,Dong:2020hxe} that the UV divergence of the coupled-channel non-relativistic effective field theory can be completely absorbed when the general form of the potential matrix is kept. The assumption made here, $V_{11}^{(s)}=0$, based on the consideration of the OZI suppression, is still valid for a reasonable variation of the scale since the two-point scalar loop integral is only logarithmically divergent, as reflected by the $\log\mu$ term in Eq.~\eqref{eq:G}.}  
Thus, in this work we have fixed $a_1(\mu)=-2.77$, with $\mu=1$~GeV, to match the $G(s)$ function at the $\jp \pi$ threshold to that in Ref.~\cite{Albaladejo:2015lob}, regulated using a Gaussian form factor with $\Lambda=1.5$~GeV. For $a_2(\mu=1\text{ GeV})$, we have considered two different values, $-2.5$ and $-3.0$, although  only the results for $a_2(\mu)=-3.0$ will be shown in plots. The LECs obtained from fits with $a_2(\mu)=-2.5$ will be included in the uncertainties of the pole analysis and are collected in Table~\ref{tab:poles}. As commented above, an alternative way to evaluate the loop integral in Eq.~\eqref{eq:Gintegral} is to use a hard cutoff $q_{\rm max}$, see e.g. Ref.~\cite{Oller:1998hw}. The subtraction constant $a_2(\mu)$ may be estimated by comparing with the loop function obtained using a  hard cutoff ($q_\text{max}$) regularization, with a natural value for $q_\text{max}$. 
Matching the loop function using the two regularizations evaluated at the $D\bar{D}^*$ threshold, $a_2(\mu =1~{\rm GeV})=-2.5$ and $-3.0$ correspond to $q_\text{max}=1.2$ and 1.8~GeV, respectively.

In summary, we have only three\footnote{Two for the case of  constant potentials.} ($C_Z$, $C_{12}$, $b$)  free parameters for the $T$-matrix, four normalization factors ($\mathcal{N}_1$ and $\mathcal{N}_1^\prime$ for $e^+e^-\to \jp\pi^+\pi^-$ at 4.26 and 4.23 GeV, respectively, $\mathcal{N}_2$ for $e^+e^-\to \pi^+D^0D^{*-}$ at 4.26 GeV, and a common one $\mathcal{N}_s$ for the $e^+e^-\to \psi(4660)\to K^+(D^{*0}D_s^-+D^0D_s^{*-})$ reaction for all $e^+e^-$ c.m. energies), three background parameters (see Eq.~\eqref{eq:bkg}) for each of the  $e^+e^-\to Y(4230)\to \jp\pi^+\pi^-$,  $e^+e^-\to Y(4260)\to \jp\pi^+\pi^-$ and $e^+e^-\to Y(4260)\to \pi^+D^{*-}D^0$, and three additional  parameters $\alpha, \beta$ and $r$ related to the point-like production of $\jp\pi^+\pi^-$, $\pi^+D^0\bar{D}^{*}$ and $K^+(D\bar{D}_s^{(*)}+D^*\bar{D}_s)$, respectively. There is a total of 19 real (or 18 in the case of constant potentials) fit parameters, to be determined from fitting to more than 220 data points. Adding a relative phase between the TS and the point-like production mechanisms barely improves the fits.

In this work, we will consider four different fit schemes, which correspond to considering or neglecting the strength ($h_S$) of the $S$-wave  $D_1 D^*\pi$ vertex  and the energy-dependent term of the $\tilde{V}^{(s)}_{22}$ diagonal part of the interaction between the two heavy-light mesons (parameter $b$ in Eq.~\eqref{eq:V22}), 
\begin{itemize}
\item Scheme IA: We use only the $D$-wave $D_1 D^*\pi$ coupling  ($h_S=0$) and take constant potentials ($b=0$).
\item Scheme IB: We extend  the  Scheme IA by considering also the energy-dependent part of the two heavy-light meson interaction ($b\neq 0$). 
\item Scheme IIA: We use both $S$- and $D$-wave $D_1 D^*\pi$ couplings  ($h_S\neq 0$) and take constant potentials ($b=0$).
\item Scheme IIB:  We extend  the Scheme IIA by considering also the energy-dependent part of the two heavy-light meson interaction ($b\neq 0$). 
\end{itemize}

As mentioned above, for each scheme two different values for the subtraction constant, $a_2(\mu= 1~{\rm GeV})=-2.5$ and $-3.0$ are considered. 
Results of the four fit scenarios are collected in Table~\ref{tab:fits}, where only the parameters of the two-meson $T$-matrix are compiled. The results obtained for constant potentials ($b=0$), {\it i.e.}, in Schemes IA and IIA, are shown in Figs.~\ref{fig:fit_A_zc} and \ref{fig:fit_A_zcs}, for the $Z_c$ and $Z_{cs}$ structures, respectively. The two fits lead to similar $\chi^2/{\rm dof}$, and the best fit curves are barely different, in particular, for the $J/\psi\pi$ invariant mass distributions. However, they correspond to distinctly different backgrounds [\textit{cf.} Eq.~\eqref{eq:bkg}], especially for the $Y(4230)\to \jp\pi\pi$ reaction. The difference can be traced back to that the $D$-wave $D_1 D^*\pi$ coupling would induce $D$-wave pions, which lead to a different Dalitz plot projection onto the $J/\psi\pi$ invariant mass distribution compared with the $S$-wave ones; see Ref.~\cite{Guo:2020oqk}. Nonetheless, the scattering $T$-matrix is well constrained.

As found in  Ref.~\cite{Albaladejo:2015lob} and shown here in Table~\ref{tab:fits}, the fit quality is significantly improved once the strength ($b$) of the  energy-dependent potential term is allowed to vary (Schemes IB and IIB).  The spectra found  for these two fits  
are shown now in Figs.~\ref{fig:fit_B_zc} and \ref{fig:fit_B_zcs} for the hidden-charm and hidden-charm strange sectors, respectively.  Both fits provide similarly quite good descriptions of the experimental distributions, with only tiny differences. As in the case of Schemes IA and IIA, displayed Figs.~\ref{fig:fit_A_zc} and \ref{fig:fit_A_zcs}, the inclusion of the $S$-wave $D_1D^*\pi$ coupling modifies the relative weight of the background in the non-strange hidden charm reactions.

A full discussion on the spectroscopic content of our fits will be given below, but we can anticipate here our main conclusion.
We see that the BESIII data sets related to the $\zc/Z_c(3885)$ and $\zcs$ signatures can be simultaneously reproduced using  SU(3) flavour symmetry, which supports the assumption that the $\zcs$ is the strange partner of the $\zc/Z_c(3885)$. {Yet, the $Z_{cs}$ structure is only significant in the data set at $M=4.681$~GeV, likely due to the enhancement induced by the triangle singularity from the diagram (3b) in Fig.~\ref{fig:feyn_zcs}, as already discussed in Refs.~\cite{Yang:2020nrt,Baru:2021ddn}. Data with higher statistics will be highly valuable. }

\begin{table*}[tb!]
\caption{Parameters of the $T$-matrix obtained for the different fit schemes discussed in the text, together with the corresponding $\chi^2/{\rm dof}$. The asterisk marks an input (fixed) value.  Here only the statistical uncertainties are presented.}
\begin{tabular}{ l | c| c| c | c c c}
\hline\hline
Scheme & $D_1D^*\pi$ coupling&  $a_2(\mu)^*$ & $\chi^2/{\rm d.o.f.}$ & $C_{12}$ [fm$^2$] & $C_Z$ [fm$^2$] & $b$ [fm$^3$]    \\
\hline
\multirow{2}{*}{\phantom{I}IA} & \multirow{2}{*}{$D$-wave} &  $-2.5 $ & 1.62 &~$0.005\pm 0.001$ &~$-0.226\pm 0.010$ & $0^*$ \\
\cline{3-7} 
& & $-3.0$ & 1.62 & ~$0.005\pm 0.001$ &~$-0.177\pm 0.006$ & $0^*$\\
\hline
\multirow{2}{*}{IIA} & \multirow{2}{*}{$S$+$D$-wave} & $-2.5$ & 1.83 &~$0.006\pm 0.001$ & ~$-0.217\pm 0.010$ & $0^* $\\
\cline{3-7}
& & $-3.0$ & 1.83 & ~$0.006\pm 0.001$ &~$-0.171\pm 0.006$ & $0^*$ \\
\hline
\multirow{2}{*}{\phantom{I}IB} &\multirow{2}{*}{$D$-wave} & $-2.5$ & $1.24$ &~$0.007\pm 0.004$ &~$-0.222\pm0.006$ & ~$-0.447\pm 0.044$\\
\cline{3-7}
& &  $-3.0$ & $1.21$ &~$0.008\pm0.001$ & ~$-0.177\pm0.004$ &~$-0.255\pm0.030$ \\
\hline
\multirow{2}{*}{IIB} & \multirow{2}{*}{$S$+$D$-wave} & $-2.5$ & 1.37 & ~$0.005\pm 0.001$ &~$-0.203\pm0.007$ & ~$-0.473\pm 0.045$ \\
\cline{3-7}
& & $-3.0$ & 1.27 & ~$0.005\pm 0.001$ &~$-0.171\pm0.005$ & ~$-0.270\pm 0.030$  \\
\hline
\hline
\end{tabular}
\label{tab:fits}
\end{table*}

\begin{figure*}[tb!]
\centering
\includegraphics[width=0.45\textwidth,height=0.28\textheight]{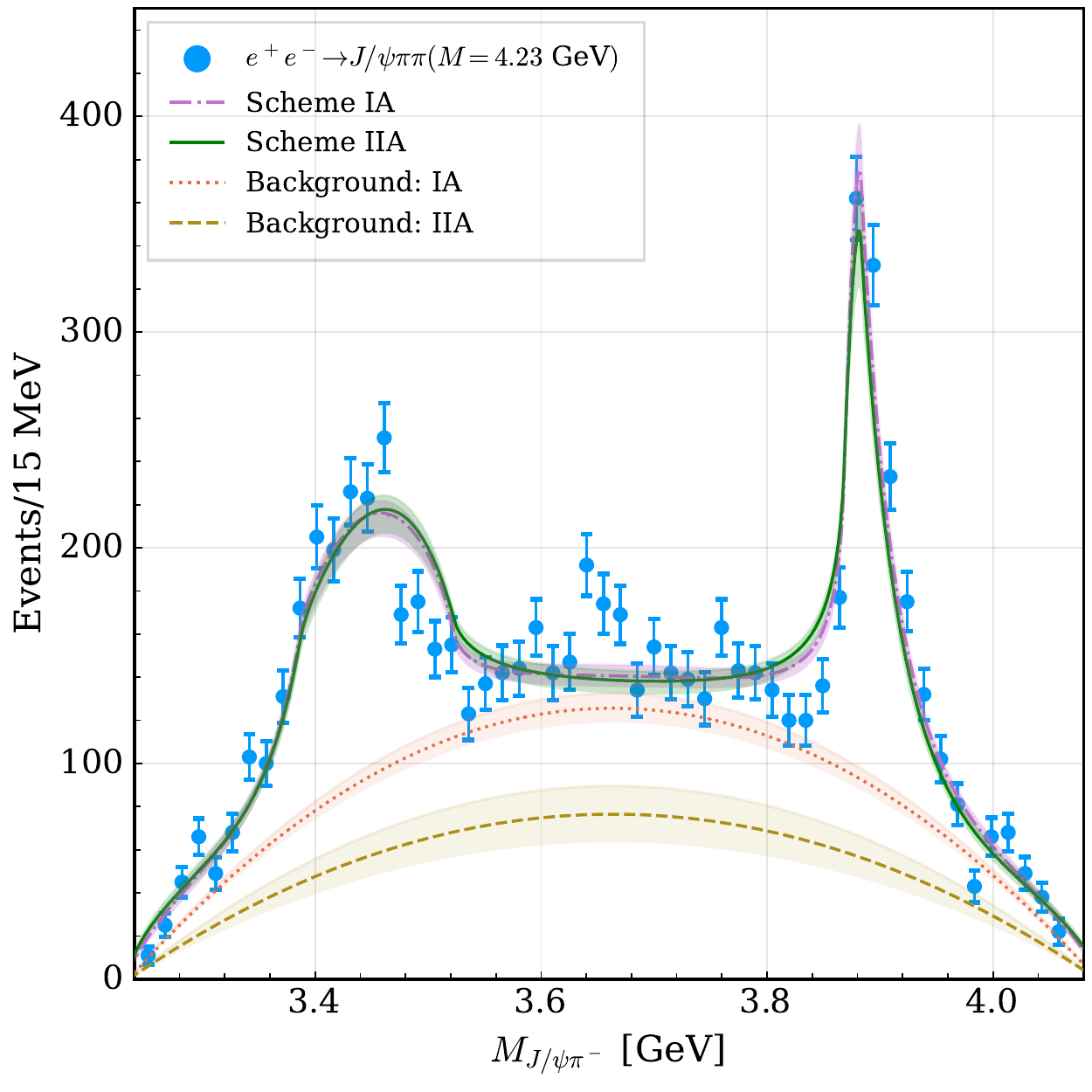}\includegraphics[width=0.45\textwidth,height=0.28\textheight]{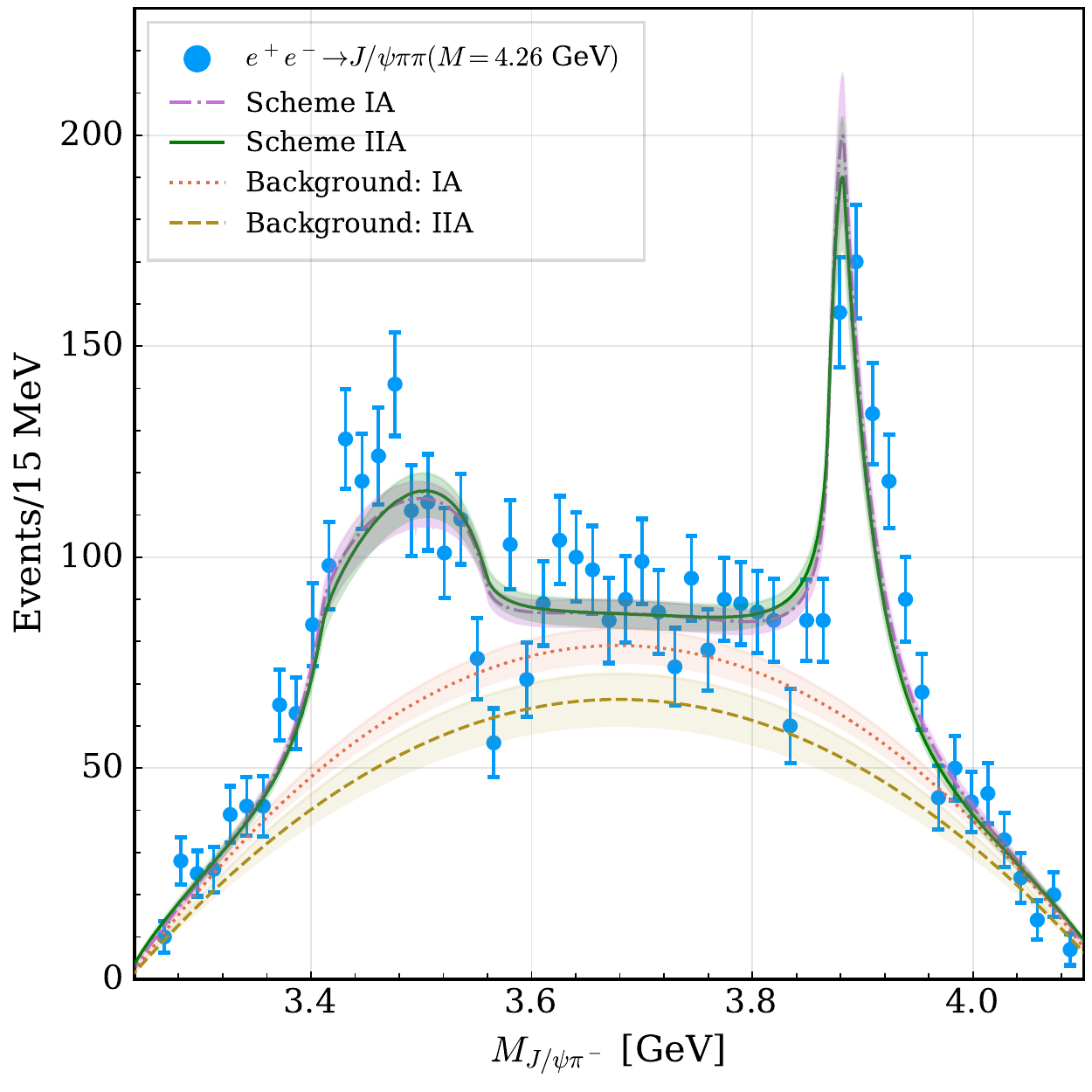}
\includegraphics[width=0.54\textwidth,height=0.24\textheight]{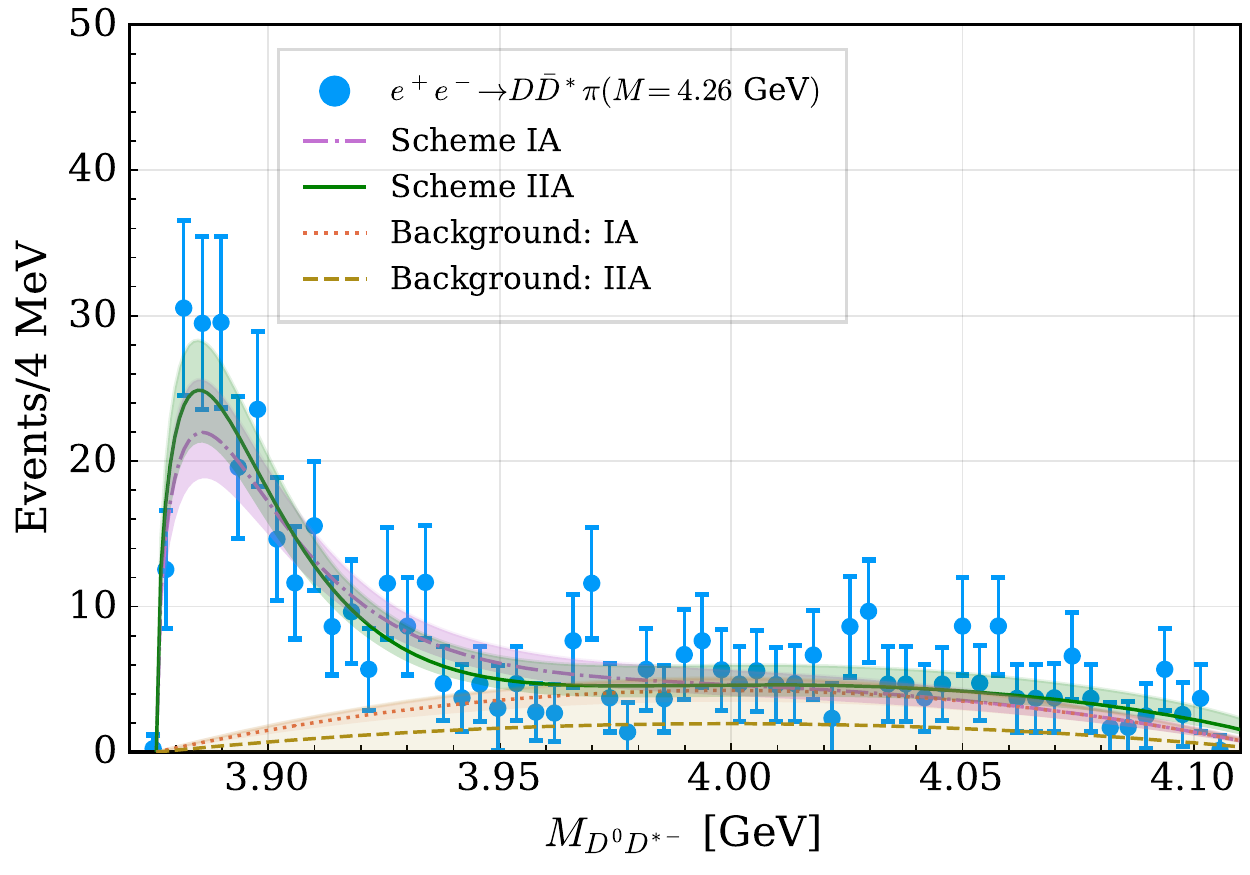}
\caption{Fitted $\jp\pi^-$ invariant mass distributions for  $e^+e^-\to \jp\pi^+\pi^-$ at the $e^+e^-$ c.m. energies $M=4.23$~GeV (upper left) and 4.26 GeV (upper right)~\cite{BESIII:2017bua}, and the  $D^0D^{*-}$ invariant mass spectrum~\cite{BESIII:2015pqw} for the $e^+e^-\to \pi^+D^0D^{*-}$ reaction at 4.26~GeV (lower panel). The results shown are  for  Schemes IA and IIA (constant potentials, {\it i.e.}, $b=0$) with the subtraction constant $a_2(\mu)$ setting to $-3.0$. The error bands (as well as for those in Figs.~\ref{fig:fit_A_zc}-\ref{fig:fit_B_zcs}) are statistical, propagated from the uncertainties quoted in Table~\ref{tab:fits}; so are the bands in the following plots. The energy resolution in the BESIII measurement is very high~\cite{BESIII:2017bua}, much finer than the bin size, and can be neglected. Thus, the effects due to the finite 15 MeV bin size  for the $e^+e^-\to J/\psi\pi\pi$ reaction (upper panels) are accounted for by averaging over the energies of each bin.}
\label{fig:fit_A_zc}
\end{figure*}

\begin{figure*}[htb!]
\centering
\includegraphics[width=1.0\textwidth,height=0.23\textheight]{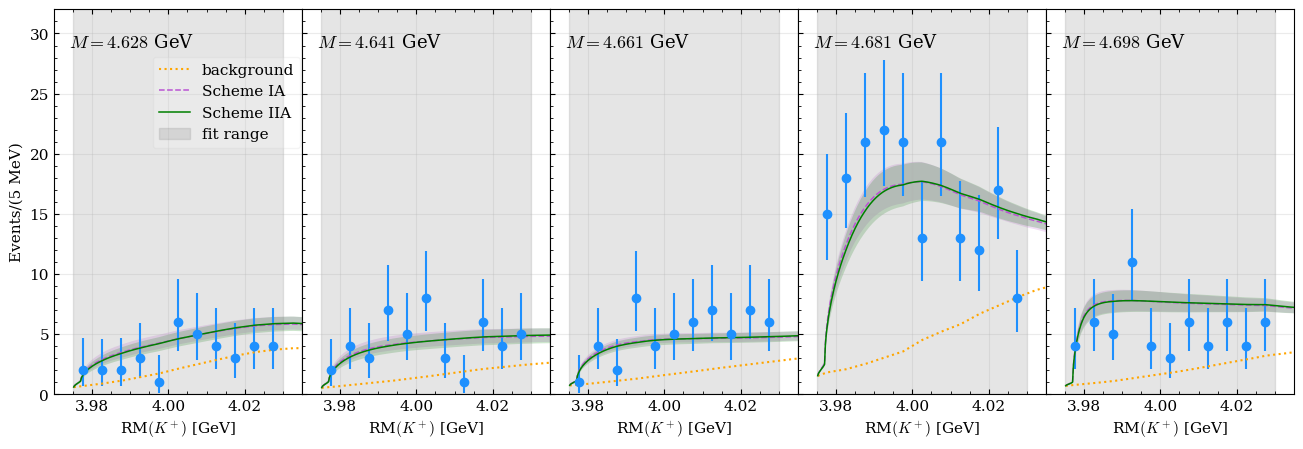}
\caption{Fitted $K^+$ recoil-mass distributions for the $e^+e^-\to K^+(D^{*0}D_s^-+D^0D_s^{*-})$ reaction~\cite{BESIII:2020qkh} at  different $e^+e^-$ c.m. energies $M=4.628$, $4.641$, $4.661$, $4.681$, and $4.698$~GeV (panels from left to right). The fitted energy region is shaded. The results shown are  for  schemes IA and IIA (constant potentials, {\it i.e.}, $b=0$) with the subtraction constant $a_2(\mu)$ setting to $-3.0$. Note that the two fit scenarios provide nearly identical predictions, which are hardly distinguishable.}
\label{fig:fit_A_zcs}
\end{figure*}

\begin{figure*}[htb!]
\centering
\includegraphics[width=0.45\textwidth,height=0.28\textheight]{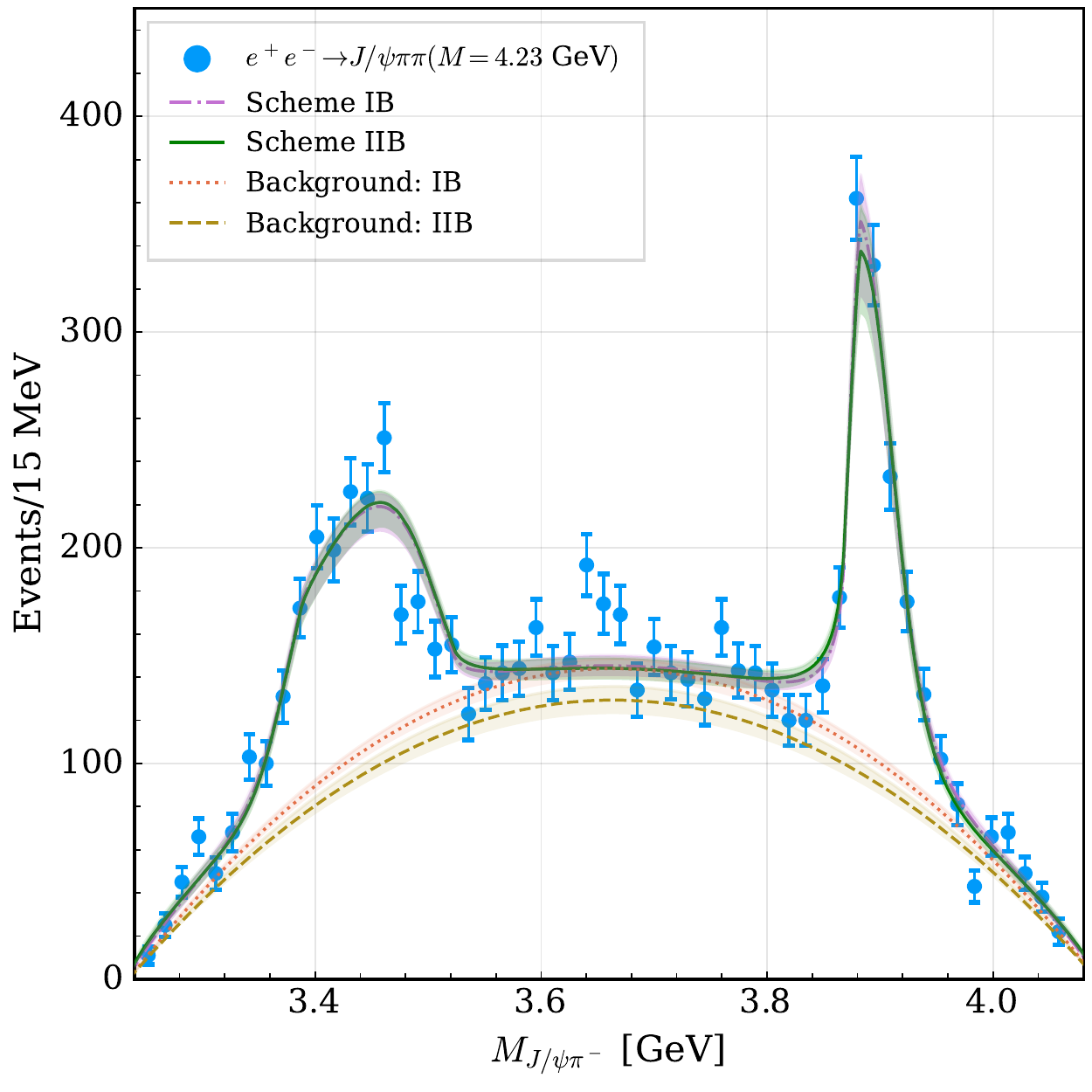}\includegraphics[width=0.45\textwidth,height=0.28\textheight]{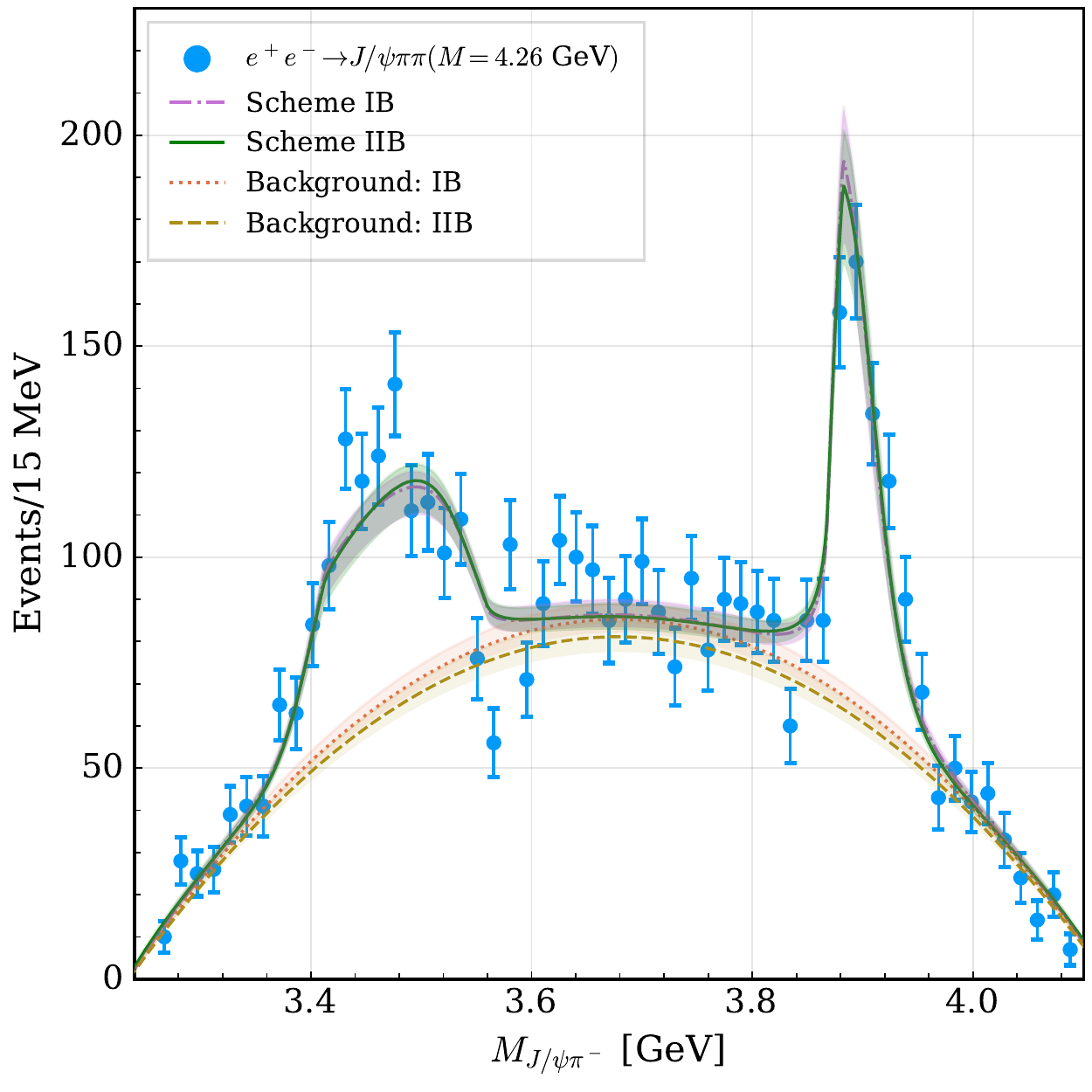}
\includegraphics[width=0.54\textwidth,height=0.24\textheight]{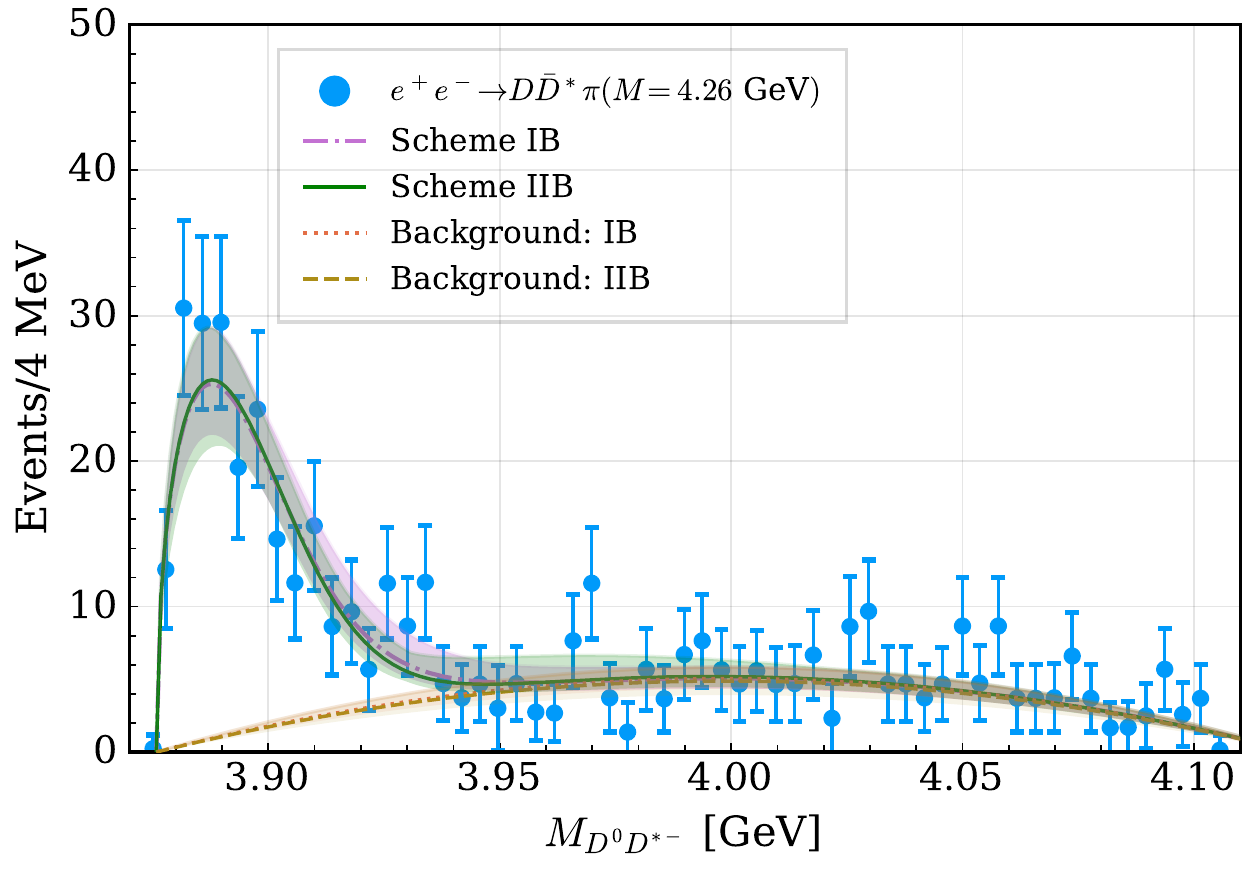}
\caption{Same as in Fig.~\ref{fig:fit_A_zc}, but for Schemes IB and IIB, which incorporate a non-vanishing energy-dependent part of the potential ($b\neq 0$).  }
\label{fig:fit_B_zc}
\end{figure*}

\begin{figure*}[htb!]
\centering
\includegraphics[width=1.0\textwidth,height=0.23\textheight]{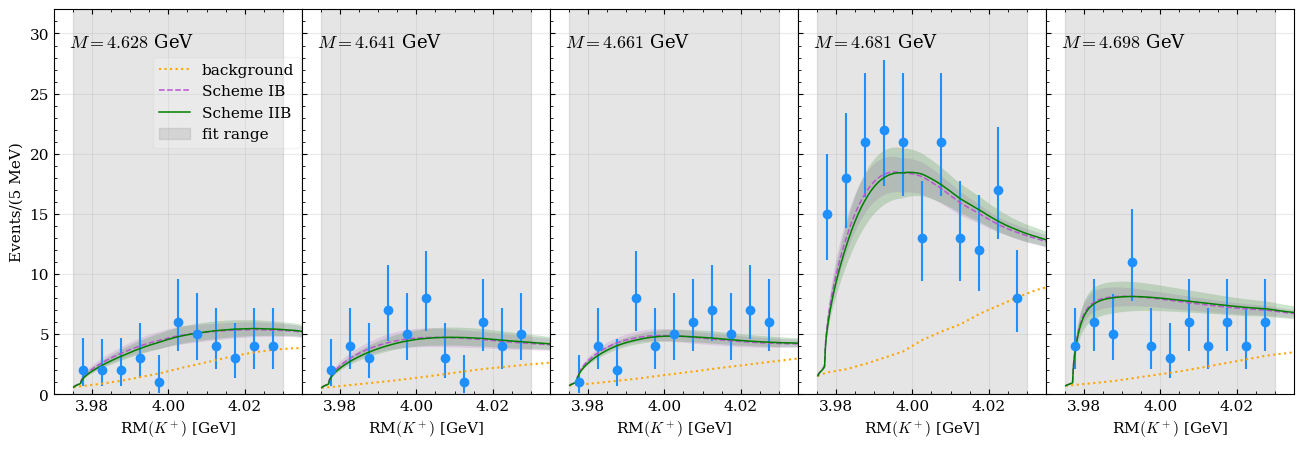}
\caption{Same as in Fig.~\ref{fig:fit_A_zcs}, but for Schemes IB and IIB, which incorporate a non-vanishing energy-dependent part of the potential ($b\neq 0$).}
\label{fig:fit_B_zcs}
\end{figure*}

Resonances and virtual states are identified as poles  in different RSs of the $T$-matrix. For a two-channel problem,
one has a 4-sheet Riemann surface in the complex energy plane. The four RSs can be accessed through the analytical continuation of the $G_i(s)$ loop functions. The expression given in Eq.~\eqref{eq:G} stands for the physical or first RS, namely, $G_i(s) = G_i^\text{I}(s)$. The loop function in the unphysical RS, $G_i^\text{II}(s)$, which is continuously connected to $G_i^\text{I}(s)$ through the cut spanning from the threshold of the $i$th channel to infinity along the positive real-$s$ axis, is obtained by  analytic continuation as~\cite{Nieves:2001wt}
\bea
G_i^\text{II}(s) = G_i^\text{I}(s) +2 i\rho_i(s),
\eea
where $\rho_i(s) =\sigma_i(s)/{(16\pi s)}$ is the corresponding two-body phase space. Different RSs can be reached by properly choosing $G_i^\text{I}(s)$ or $G_i^\text{II}(s)$ for different channels. By denoting $G_i^\text{I}(s)$ [$G_i^\text{II}(s)$] by plus [minus], which is the sign of the imaginary part of the c.m. momentum of particles in the $i$th channel, the four RSs can be labeled as RS$_{\pm\pm}$. In this convention, the physical RS for the 2-coupled channels is labeled as RS-I = RS$_{++}$, and the other three RSs are: RS-II = RS$_{-+}$, RS-III = RS$_{--}$, and RS-IV = RS$_{+-}$. RS-II is connected to the physical region through the interval between the 1st and 2nd thresholds, and RS-III is connected to the physical region above the 2nd threshold along the real axis. RS-IV is not directly connected to the physical region, however the poles located on RS-IV  can still leave impact on the physical observables. The real and imaginary parts of a pole are identified as the mass and half width, respectively, for the corresponding resonance. 

For Schemes IB and IIB, the poles are found above the $D^*\bar D_{(s)}$ threshold on RS-III, as shown in Table~\ref{tab:poles}. They correspond to resonances which can decay to $\jp/\pi$ ($\jp \bar K$) and $D\bar{D}^*$ ($D^*\bar D_s/D\bar{D}_s^*$).\footnote{In this work, we use the sum of the masses of the $D_s$ and $\bar{D}^*$ mesons as the second threshold for the strange sector. The difference between $(m_{D_s}+m_{D^*})$ and $(m_{D_s^*}+m_D)$ is small ($< 2$ MeV) and it would produce negligible uncertainties compared to the errors of the experimental distributions.} The results for the non-strange sector are in agreement within errors  with those found in Ref.~\cite{Albaladejo:2015lob}  for the  $\Lambda_2=1.0$ GeV case.  {We postpone the comparison of the predictions found here   for the strange sector with those obtained in our previous analysis of Ref.~\cite{Yang:2020nrt} to the next section, where some SU(3) breaking effects will be included.} 

For Schemes IA and IIA ($b=0$), the description of the experimental spectra is not as good as for Schemes IB and IIB, with a larger $\chi^2/{\rm dof}$. The poles are found located below the $D^* \bar D_{(s)}$ thresholds in  RS-IV with negligible imaginary parts which are caused by the small $C_{12}$ off-diagonal couplings (see  Table~\ref{tab:fits}). 
These poles would move into the real axis on the unphysical RS of the $D^* \bar D_{(s)}$ single-channel scattering amplitude if the $\jp \pi$ ($\jp \bar K$) channel was switched off. Therefore, the poles obtained in Schemes IA and IIA are identified as $D^*\bar D_{(s)}$ virtual states, which do not correspond to {spatially localized} particles in the sense that their spatial wave functions are not normalizable (neither does that of a resonance). However, such singularities, if located near threshold, can significantly modify the line shapes at the vicinity of the $D^*\bar D_{(s)} $ threshold. While the pole positions  hardly change when the the subtraction constant $a_2(\mu)$ varies from $-2.5$ to $-3.0$, the inclusion of the $S$-wave $D_1D^*\pi$ vertex, in addition to the $D$-wave coupling, produces a bigger impact on the pole positions.  
It is interesting to notice that the real part of the virtual state pole in Schemes IA and IIA is about 100~MeV below the corresponding threshold, in line with the HAL QCD result for the $Z_c(3900)$~\cite{HALQCD:2016ofq,Ikeda:2017mee}. This occurs when the attraction is relatively weak. However, when the pole is located above threshold, as in Schemes IB and IIB, it is much closer to threshold.

\begin{table*}[tb!]
\caption{Masses and half widths of the  $\zc$, $\zcs$, $Z_c^*$, and $Z_{cs}^*$ states. The isovector $Z_{c}^*$ [isodoublet $Z_{cs}^*$] is a $D^*\bar{D}^*$  [$D^*\bar{D}_{s}^*$] molecule, with $J^{PC}=1^{+-}$ [$J^{P}=1^{+}]$, that would be the HQSS partner of the $Z_{c}$ [$Z_{cs}$]. The uncertainties presented are obtained by propagating the errors and statistical correlations of the besfit parameters.}
\begin{tabular}{ l | c | c | c c  |  cc| cc |cc}
\hline\hline
\multirow{2}{*}{Scheme}  & \multirow{2}{*}{$D_1D^*\pi$ coupling} &  \multirow{2}{*}{$a_2(\mu)$} & \multicolumn{2}{c}{$Z_c$ [MeV]} &  \multicolumn{2}{c}{$Z_{cs}$ [MeV]} & \multicolumn{2}{c}{$Z_c^*$ [MeV]} & \multicolumn{2}{c}{$Z_{cs}^*$ [MeV]}    \\
\cline{4-11}
&&& {\rm Mass} & $\Gamma/2$ & {\rm Mass} & $\Gamma/2$ & {\rm Mass} & $\Gamma/2$ & {\rm Mass} & $\Gamma/2$\\
\hline
\multirow{2}{*}{\phantom{I}IA} & \multirow{2}{*}{$D$-wave} &  $-2.5 $ & $3813^{+21}_{-28}$ & virtual & $3920^{+18}_{-26}$ & virtual & $3962^{+19}_{-25}$ & virtual & $4069^{+12}_{-16}$ & virtual \\
\cline{3-11} 
& & $-3.0$ & $3812^{+22}_{-26}$ & virtual & $3924^{+19}_{-23}$ & virtual & $3967^{+19}_{-22}$ & virtual & $4078^{+17}_{-13}$ & virtual \\
\hline
\multirow{2}{*}{IIA} & \multirow{2}{*}{$S$+$D$-wave} & $-2.5$ & $3799^{+24}_{-33}$ & virtual & $3907^{+22}_{-31}$ & virtual & $3949^{+22}_{-30}$ & virtual & $4057^{+20}_{-28}$ & virtual\\
\cline{3-11}
& & $-3.0$ & $3798^{+25}_{-31}$ & virtual  & $3911^{+17}_{-27}$ & virtual & $3955^{+22}_{-27}$ & virtual & $4067^{+19}_{-25}$ & virtual\\
\hline
\multirow{2}{*}{\phantom{I}IB} &\multirow{2}{*}{$D$-wave} & $-2.5$ & $3897^{+4}_{-4}$ & $37^{+8}_{-6}$ & $3996^{+4}_{-4}$ & $37^{+8}_{-6}$ & $4035^{+4}_{-4}$ & $37^{+8}_{-6}$ &  $4137^{+4}_{-4}$ & $36^{+7}_{-6}$ \\
\cline{3-11}
& &  $-3.0$ & $3898^{+5}_{-5}$ & $38^{+10}_{-~7}$ & $3996^{+5}_{-6}$ & $35^{+9}_{-6}$ & $4035^{+4}_{-5}$ & $34^{+9}_{-6}$ & $4136^{+5}_{-6}$ & $33^{+8}_{-6}$ \\
\hline
\multirow{2}{*}{IIB} & \multirow{2}{*}{$S$+$D$-wave} & $-2.5$ & $3902^{+6}_{-6}$ & $38^{+9}_{-6}$ & $4002^{+6}_{-6}$ & $38^{+9}_{-7}$ & $4042^{+5}_{-5}$ & $38^{+9}_{-7}$  & $4144^{+5}_{-6}$ & $37^{+9}_{-7}$ \\
\cline{3-11}
& & $-3.0$ & $3902^{+5}_{-5}$ & $37^{+9}_{-6}$ & $4000^{+5}_{-6}$ & $35^{+8}_{-7}$ & $4039^{+5}_{-6}$ & $35^{+8}_{-6}$ & $4140^{+5}_{-6}$ & $33^{+8}_{-6}$ \\
\hline
\hline
\end{tabular}
\label{tab:poles}
\end{table*}

In the heavy quark limit, the interactions of the  isovector $J^{PC}=1^{+-}$ $D^*\bar{D}^*$ and  of the isodoublet $J^P=1^+$ $D^*\bar D_s^*$ pairs are equal to those of $D\bar{D}^*$ and $\left(D\bar{D}^*_s-D^* \bar D_s\right)/ \sqrt{2}$, respectively (see Eqs.~\eqref{eq:Vdd} and \eqref{eq:Vjp}). The existence of the $Z_c$ and $Z_{cs}$ hints at the possible existence of HQSS partner resonances with the same quantum numbers. Based on the LECs obtained from the different fits compiled in Table~\ref{tab:fits}, additional $D^* \bar D_{(s)}^*$ states, denoted by $Z_{c(s)}^*$, can be predicted, provided that the $D^*\bar D_{(s)}$-$D^*\bar D_{(s)}^*$ coupled-channel effects are neglected. Analogous to the $Z_c$ and $Z_{cs}$, the $Z_c^*$ and $Z_{cs}^*$ appear to be virtual states with poles in RS-IV (of $\jp\pi$-$D^*\bar{D}^*$ or $\jp \bar K$-$D^*_s\bar{D}^*$ two-channel amplitudes) below the higher thresholds for Schemes IA and IIA, while for Schemes IB and IIB, they become resonances, above the thresholds with poles located in RS-III. The predictions for masses and half widths of the $Z_c^*$ and $Z_{cs}^*$ are collected in Table~\ref{tab:poles}.

{ 
The spin partners predicted in Table~\ref{tab:poles} were obtained by neglecting the channel coupling between the pseudoscalar-vector and vector-vector meson pairs. Yet, there is a possibility that the channel coupling might be strong.
The analysis done in Ref.~\cite{Baru:2021ddn} fits to the BESIII data of $e^+e^-\to K^+(D^{*0}D_s^-+D^0D_s^{*-})$ in the whole range  of RM$(K^+)$ considering  constant contact interactions in the $D\bar D^*_s/D^*\bar D_s$-$D^*\bar D_s^*$ coupled channels. Depending on how the interaction is organized in the coupled channels, radically different fits were obtained. 
The fits with weak $D\bar D^*_s/D^*\bar D_s$-$D^*\bar D_s^*$ channel coupling (fit~1 and fit~$1'$ therein) agree well with that in Ref.~\cite{Yang:2020nrt}. 
The pole closest to the physical region is a virtual state pole, and thus also consistent with the analysis here in schemes with $b=0$; in this case, spin partners of the $\zc$ and $\zcs$ were also predicted close to the $D^*\bar D^*$ and $D^*\bar D^*_s$ thresholds, respectively.
However, for fit~2 in Ref.~\cite{Baru:2021ddn}, the $D\bar D^*_s/D^*\bar D_s$-$D^*\bar D_s^*$ channel coupling is stronger than the diagonal interaction, and it is mainly because of the channel coupling that the $\zc$ and $\zcs$ are generated. In this case, the $\zc$ ($\zcs$) does not have a spin partner with the same $J^{PC}$ ($J^P$).
A comprehensive analysis of all data for the $\zc$, $\zcs$ and $Z_c(4020)/Z_c(4025)$ would be helpful to finally pin down the role of the channel coupling between the pseudoscalar-vector and vector-vector meson pairs.
}

It is observed in Table~\ref{tab:poles} that the mass of the $Z_c^*$ is close to that of the $Z_c(4020/4025)$, suggesting that this latter (observed) resonance is the HQSS partner of the $Z_c(3900)$. 
Although the  $Z_c^*$ width is larger than that of the $Z_c(4020/4025)$ reported by BESIII, a direct comparison is not very meaningful since the BESIII analysis was made using a Breit-Wigner parametrization. 
Also, as mentioned in Ref.~\cite{Yang:2020nrt},  just before the $Z_{cs}(4220)$ peak there is a dip around the $ D^*\bar D_s^*$ threshold, and it is well possible that the dip is a consequence of the $Z_{cs}^*$ (for a general discussion of the appearance of a dip-like near-threshold structure, see Ref.~\cite{Dong:2020hxe}).
It is plausible that the  $Z_{cs}(4000)$ reported by LHCb~\cite{LHCb:2021uow} has the same origin as the pole part of the $Z_{cs}(3985)$ signal reported by BESIII despite that the reported width of the former is much larger than that of the latter. In order to reach a firm conclusion on this, it would be very valuable to conduct in the future a joint analysis of both data sets within the framework derived here.
Nevertheless, the above claim, proposed in the noted added of Ref.~\cite{Yang:2020nrt}, finds support in  Ref.~\cite{Ortega:2021enc}, where both the BESIII and LHCb data can be described well within the same model, and with the $Z_{cs}$ and $Z_{cs}^*$ appearing as  virtual states.

Finally, we briefly discuss the production of the $Z_{c(s)}$. In Eq.~\eqref{eq:ampA1}, the point-like production of the $J/\psi\pi\pi$ (with $\pi\pi$ in $D$-wave) is modeled by a constant coupling $\alpha$, and the final state interaction for $J/\psi\pi\to J/\psi \pi$ direct transition is neglected, since it is OZI suppressed. Therefore, the point-like production can not generate a peak  structure in the $J/\psi \pi$ mass distributions around the $D\bar{D}^*$ threshold. For the $D\bar{D}\pi$ final state, all the fit schemes produce negligible values of $\beta$  which have magnitudes much smaller than their uncertainties. For the $e^+e^-\to K^+(D^{*0}D_s^-+D^0D_s^{*-})$ reaction, the point-like production contribution for the  $M=4.628$ and $4.641$ GeV $e^+e^-$ c.m. energies are comparable with that of the TS mechanism. For  $M=4.661$, $4.681$ and $4.698$ GeV, which are the energies closer to the TS in the $M$ variable, are dominated by the TS mechanism.

\section{SU(3) flavor symmetry breaking effects}\label{sec:su3b}

Up to now, we have employed two-meson potentials in the SU(3) limit, which allows to use common LECs ($C_Z$, $C_{12}$ and $b$) for the $D\bar{D}^*$ and $ D^*\bar D_s/D\bar D^*_s $ interactions, cf. Eq.~\eqref{eq:V22}. However, the SU (3) flavor is not an exact symmetry and we could expect violations of around $20\%$.

In addition, we should note that the parameters related to the $T$-matrix are mainly determined by the $Z_c$ related mass distributions, since they incorporate more data points with smaller uncertainties, which strongly  constrain the fits. Although both the $Z_c$ and $Z_{cs}$,  can be well described in the SU(3) limit, as demonstrated in Sec.~\ref{sec:fit}, cf. Figs.~\ref{fig:fit_B_zc} and \ref{fig:fit_B_zcs}, we observe that the description of the $Z_{cs}$ structure at $M=4.681$ GeV can be probably improved by including SU(3) breaking effects. In Table~\ref{tab:poles}, we observe that the mass differences between the $Z_{cs}$ and $Z_c$ states in Schemes IB and IIB are 98--99~MeV, evaluated using the central values, which are consistent with the predictions of Ref.~\cite{Yang:2020nrt} in the case of the SU(3) limit, {\it i.e.}, 100--102~MeV evaluated using the central values obtained with different UV cutoffs. To investigate effects from light-flavor symmetry violations, we consider two ways to include SU(3) breaking terms: 
\begin{itemize}
\item (b): introducing a breaking term in the diagonal $\left(D\bar{D}^*_s-D^* \bar D_s\right)/ \sqrt{2}$ potential,  $C_Z^s = C_Z+\delta_Z$, and keeping the other terms of  the two-body interactions unchanged.
\item (c):  introducing a breaking term in  the off-diagonal $\jp \bar K$-$\left(D\bar{D}^*_s-D^* \bar D_s\right)/ \sqrt{2}$ potential, $C_{12}^s=C_{12}+\delta_{12}$, and keeping the other terms of  the two-body interactions unchanged.
\end{itemize}
An SU(3) breaking correction in the energy-dependent term ($b$-term) is not considered because this LEC is of higher order in the heavy-meson momentum expansion. On the other hand, as shown in the previous section, the pole positions are insensitive to the value of the subtraction constant $a_2(\mu)$, thus in this section we set $a_2(\mu)=-3.0$ for definiteness. In Table~\ref{tab:para_su3b} we show the results of these fits, and the poles so obtained are shown in Table~\ref{tab:pole_su3b}.

The line shapes for the $Z_c$ structure are almost unchanged after introducing the SU(3) breaking terms. The description of the ${\rm RM}(K^+)$  distributions, especially  for $M=4.681$ GeV, is significantly improved in schemes of type (b). However, for schemes of type (c), the description of the ${\rm RM}(K^+)$ distributions has been less improved (see Table~\ref{tab:para_su3b}). Therefore and to illustrate the effects, we   show  the comparison between the results obtained in the SU(3) limit and the type  (b) SU(3) breaking scenario  in Fig.~\ref{fig:su3b}. In this figure, we consider a full $S+D$-wave $D_1D^*\pi$ vertex. 

While the introduction of $\delta_{12}$ for the strange sector does barely improve the fit quality, it, however, leads to significantly larger uncertainties for the $Z_{cs}^{(*)}$ parameters, see those for the schemes of type (c) in Table~\ref{tab:pole_su3b}. For the schemes of type (b), the decreased value of $\chi^2$/dof is mainly caused by the improved description of the $Z_{cs}$ structure at $M=4.681$ GeV. It is easy to conclude from  Fig.~\ref{fig:su3b} that the {$Z_{cs}$} pole position approaches to the $D^*\bar D_s$ threshold,  as can be confirmed in Table~\ref{tab:pole_su3b}, which produces a more pronounced enhancement of the ${\rm RM}(K^+)$ distribution in the region close to threshold. 
Let us focus on Schemes IB and IIB with SU(3) breaking of type (b), which lead to  smaller $\chi^2/{\rm dof}$ than Schemes IA and IIA. The mass and width of $Z_c$ are almost unchanged compared to those found in the SU(3) limit, while  the mass and width of the $Z_{cs}$ resonance are reduced by around 15 MeV and 10 MeV, respectively. For convenience, in Fig.~\ref{fig:allpoles} we collect all the resonance poles in Tables~\ref{tab:poles}~and~\ref{tab:pole_su3b}. {The results of this work, when SU(3) breaking effects are considered, compare reasonably well with those obtained in Ref.~\cite{Yang:2020nrt} from straight fits to the BESIII ${\rm RM}(K^+)$ distributions. }

With the parameters in Table~\ref{tab:para_su3b}, we estimate the size of the SU(3) breaking terms as
\bea
\Delta_{\rm b} = \frac{\delta_Z}{C_Z+\delta_Z/2}\sim  20\% ,
\eea
which is consistent with a naive expectation. We stress  that the $\Delta_{\rm b} $ value is not the unique evaluation of the level of SU(3) violation since the LECs $C_Z$ and $\delta_Z$ are not physical observables and they are renormalization scale-dependent. However, this quantity still provides a rough assessment of the SU(3) violation.

\begin{table*}[tb!]
\caption{Parameters of the $T$-matrix obtained for the different fit schemes which include SU(3) breaking effects (see text for details), together with the corresponding $\chi^2/{\rm dof}$. The asterisk  marks an input (fixed) value. Here we have set $a_2(\mu)=-3.0$. Only the statistical uncertainties are presented.}
\begin{tabular}{  l | c | c c c c c}
\hline\hline
 Scheme & $\chi^2/{\rm dof}$ & $C_{12}$ [fm$^2$] & $C_Z$ [fm$^2$]  & $\delta_Z$ [fm$^2$] & $\delta_{12}$ [fm$^2$] & $b$ [fm$^3$] \\
\hline
  \phantom{I}IA (b) & 1.58 &  ~$0.004\pm 0.001$ & ~$-0.171\pm 0.006$ & $-0.038\pm 0.009$ & - & $0^*$ \\
\hline
  \phantom{I}IA (c) & 1.62  &   ~$0.004\pm 0.003$ &  ~$-0.175\pm 0.007$ & - & $0.230\pm0.110$ & $0^*$\\
\hline
 IIA (b) & 1.80 &  ~$0.006\pm 0.001$ &  ~$-0.166\pm 0.007$ & $-0.042\pm 0.007$ & - & $0^* $\\
\hline
 IIA (c) &  1.84 &  ~$0.005\pm 0.001$ &  ~$-0.170\pm 0.006$ & - & $0.253\pm0.103$  & $0^*$ \\
\hline
\phantom{I}IB (b)  & $1.19$ &   ~$0.007\pm 0.001$ &  ~$-0.175\pm0.004$ & $-0.022\pm 0.009$ & - & $-0.251\pm 0.030$\\
\hline
 \phantom{I}IB (c) & $1.21$ &  ~$0.006\pm0.001$ &  ~$-0.177\pm0.004$ & - & $0.194\pm0.101$ &  $-0.255\pm0.030$ \\
\hline
 IIB (b) & 1.24 &  ~$0.005\pm 0.001$ &  ~$-0.167\pm0.005$ & $-0.029\pm 0.009$ & - & $-0.265\pm 0.029$ \\
\hline
  IIB (c) & 1.26 &  ~$0.005\pm 0.001$ &  ~$-0.169\pm0.005$ & - & $0.235\pm 0.093$ &$-0.270\pm 0.030$  \\
\hline
\hline
\end{tabular}
\label{tab:para_su3b}
\end{table*}

\begin{figure*}[htb!]
\centering
\includegraphics[width=0.495\textwidth]{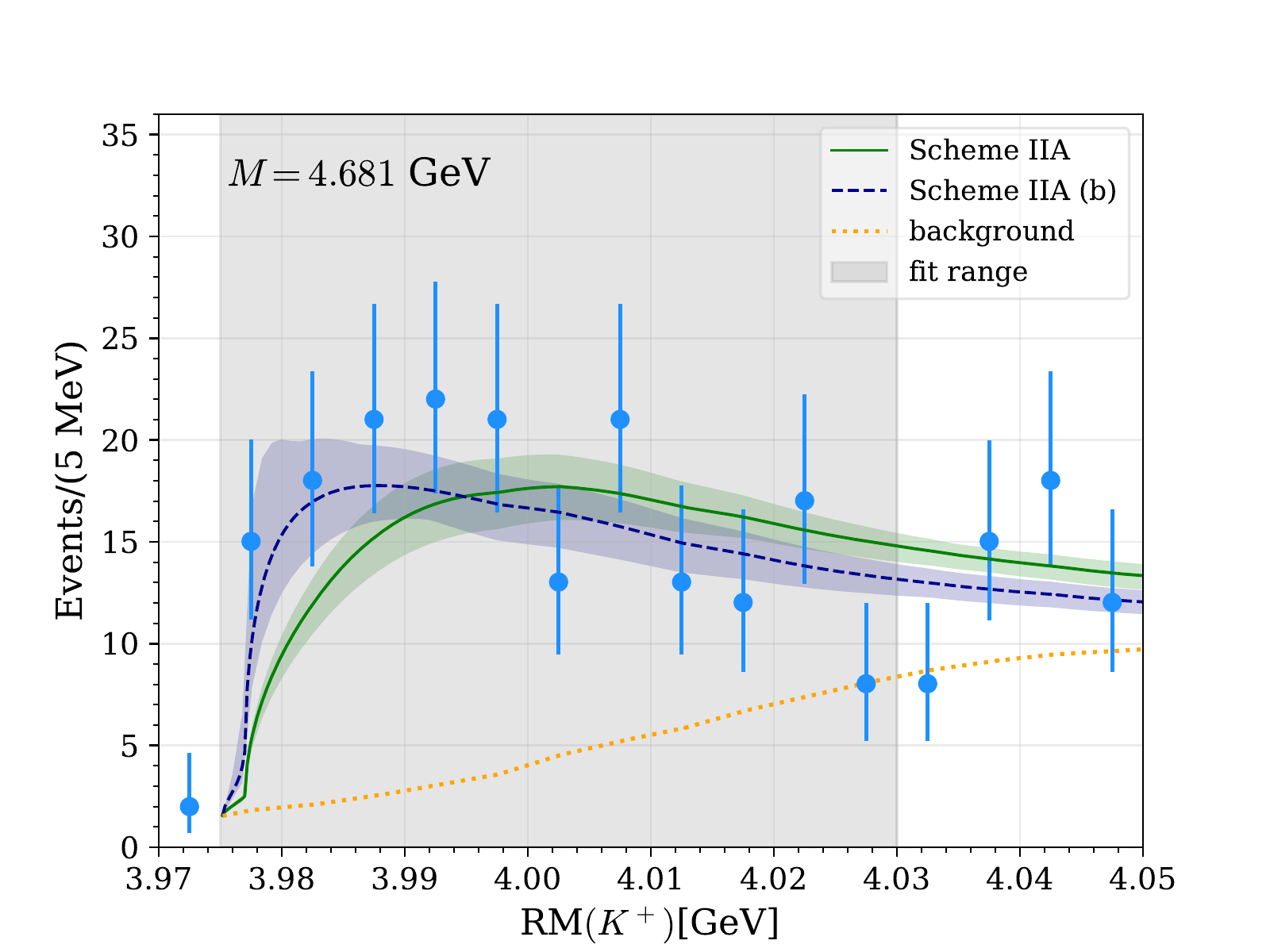}
\includegraphics[width=0.495\textwidth]{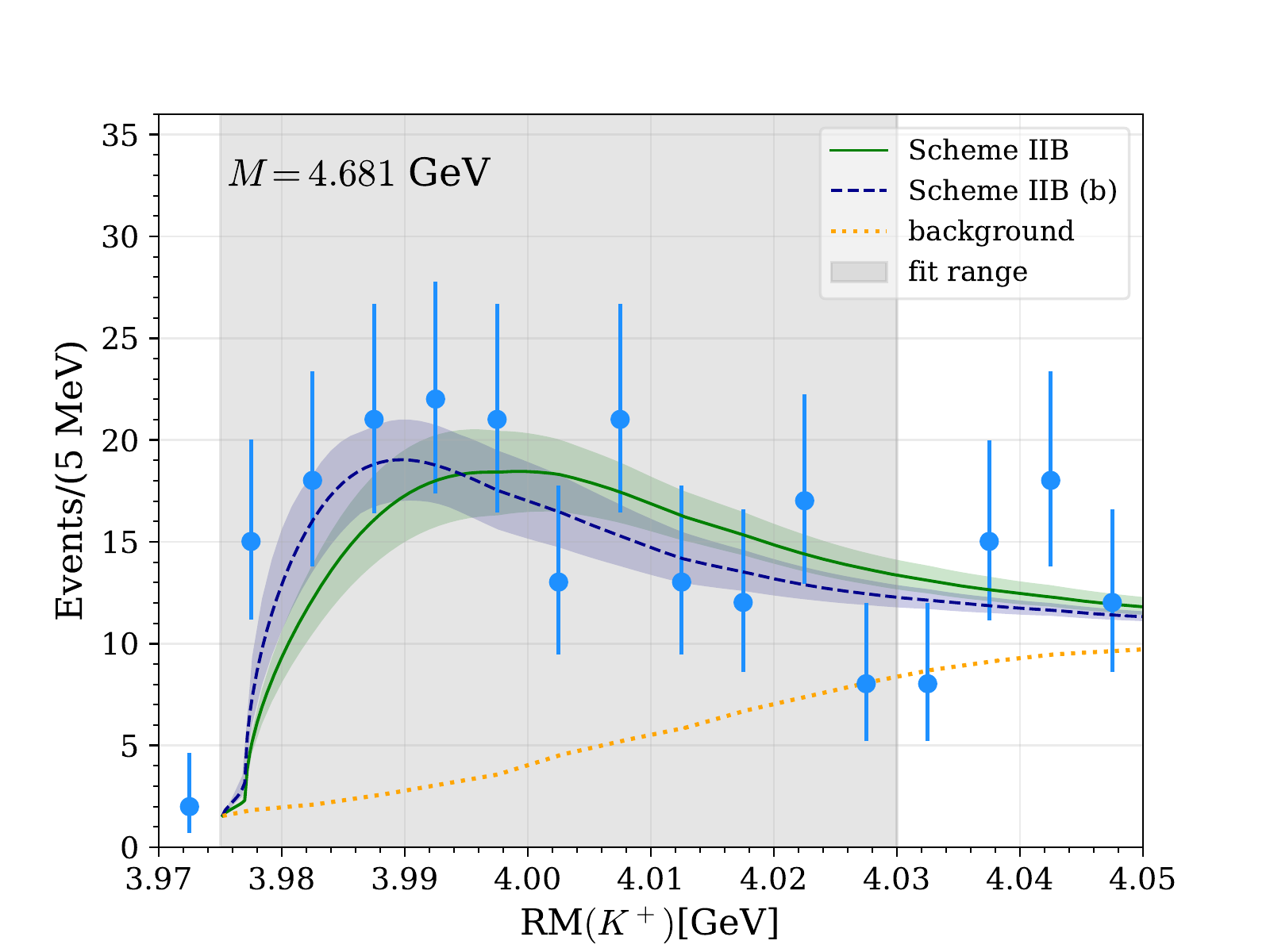}
\caption{Comparison of the ${\rm RM}(K^+)$ distributions at $M=4.681$ GeV  obtained in the SU(3) limit  and the SU(3) breaking scenario of type (b). The fitted energy region is shaded. The energy-dependent part of potential, controlled by the LEC $b$ is  zero (different to zero and fitted to data) in the left (right) panel. In all cases, the full $S+D$-wave $D_1D^*\pi$ vertex is used. The error bands are statistical, propagated from the uncertainties quoted in Table~\ref{tab:para_su3b}.}
\label{fig:su3b}
\end{figure*}

\begin{table*}[htb!]
\caption{Masses and half widths of the  $\zc$, $\zcs$, $Z_c^*$, and $Z_{cs}^*$ resonances obtained with different fit schemes, which include SU(3) breaking effects (see text for details). The subtraction constant $a_2(\mu)$ is fixed to $-3.0$. {Uncertainties on the resonance parameters are derived from the errors and statistical correlations of the best fit parameters.}}
\begin{tabular}{ l |  c c  |  cc| cc |cc}
\hline\hline
\multirow{2}{*}{Scheme}  &   \multicolumn{2}{c}{$Z_c$ [MeV]} &  \multicolumn{2}{c}{$Z_{cs}$ [MeV]} & \multicolumn{2}{c}{$Z_c^*$ [MeV]} & \multicolumn{2}{c}{$Z_{cs}^*$ [MeV]}    \\
\cline{2-9}
& {\rm Mass} & $\Gamma/2$ & {\rm Mass} & $\Gamma/2$ & {\rm Mass} & $\Gamma/2$ & {\rm Mass} & $\Gamma/2$\\
\hline
  \phantom{I}IA (b) & ~$3796^{+26}_{-30}$ & virtual & $3967^{+7}_{-10}$ & virtual & ~$3954^{+23}_{-27}$ & virtual & $4114^{+5}_{-7}$ & virtual \\
\hline
  \phantom{I}IA (c) & ~$3808^{+20}_{-27}$ & virtual  & ~$3948^{+55}_{-35}$ & virtual & ~$3964^{+17}_{-24}$ & virtual & ~$4101^{+59}_{-32}$ & virtual \\
\hline
 IIA (b) & ~$3784^{+20}_{-19}$ & virtual & ~$3967^{+\phantom{1}8}_{-12}$ & virtual & ~$3943^{+17}_{-17}$ & virtual & $4114^{+5}_{-8}$ & virtual \\
\hline
 IIA (c) &  ~$3794^{+24}_{-32}$ &  virtual & ~$3946^{+65}_{-37}$ & virtual & ~$3952^{+21}_{-28}$ & virtual &  ~$4100^{+61}_{-35}$ & virtual  \\
\hline
\phantom{I}IB (b)  & $3900^{+4}_{-5}$ & $39^{+9}_{-7}$ & $3982^{+9}_{-12}$ & $27^{+8}_{-9}$ & $4036^{+4}_{-5}$ & $37^{+8}_{-6}$ & $4122^{+10}_{-11}$ & $24^{+8}_{-11}$ \\
\hline
 \phantom{I}IB (c) & $3899^{+4}_{-5}$ & $38^{+10}_{-\phantom{1}6}$ & ~$3985^{+12}_{-36}$ & $40^{+18}_{-\phantom{1}8}$ & $4035^{+4}_{-5}$ & $35^{+9}_{-7}$ & ~$4124^{+13}_{-36}$ & $39^{+25}_{-10}$ \\
\hline
 IIB (b) & $3904^{+5}_{-5}$ & $39^{+9}_{-7}$ & $3984^{+\phantom{1}9}_{-10}$ & $26^{+8}_{-8}$ & $4041^{+5}_{-5}$ & $37^{+8}_{-6}$ & ~$4124^{+\phantom{1}9}_{-11}$ & $23^{+8}_{-9}$  \\
\hline
  IIB (c) & $3903^{+5}_{-5}$ & $38^{+9}_{-6}$ &  ~$3986^{+14}_{-27}$ & $42^{+16}_{-10}$ & $4040^{+5}_{-5}$ & $35^{+8}_{-6}$ & ~$4125^{+15}_{-27}$ & $42^{+19}_{-11}$ \\
\hline
\hline
\end{tabular}
\label{tab:pole_su3b}
\end{table*}

\begin{figure*}
\includegraphics[height=6cm,keepaspectratio]{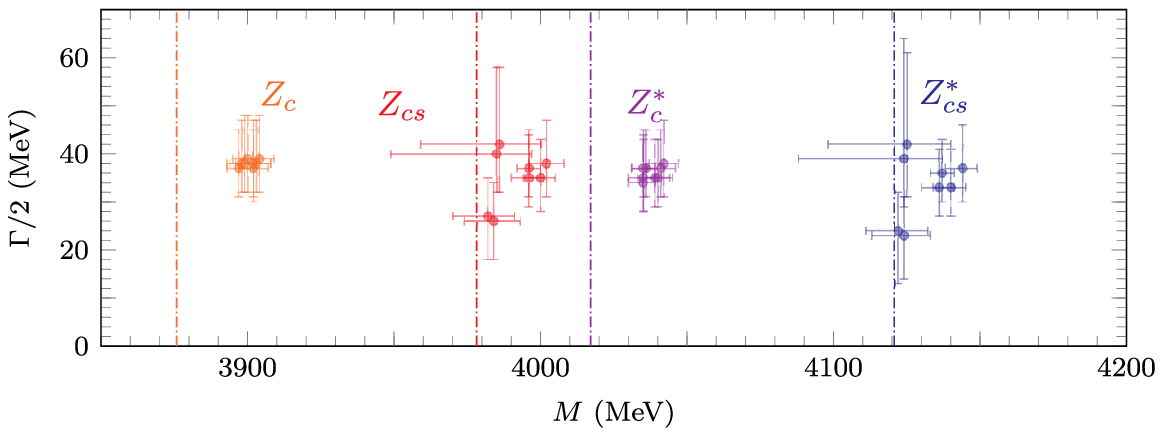}
\caption{Resonance pole parameters, collected from Tables~\ref{tab:poles}~and~\ref{tab:pole_su3b},  obtained in different fits studied in this work, with an energy-dependent term in the diagonal $D^{(*)}\bar D_{(s)}^{(*)}$ interaction. The respective nearby thresholds are also shown. \label{fig:allpoles}}
\end{figure*}

\section{Conclusions}\label{sec:con}

We have performed a combined analysis of the BESIII data for both the $\zc$ and $\zcs$ structures, and it is found that the data can be well described assuming that the latter is an SU(3) flavor partner of the former one. We have improved on the previous analysis of Ref.~\cite{Albaladejo:2015lob} by computing the amplitude for the $D_1\bar{D}D^*$ triangle diagram  considering  both  $D$- and $S$-wave $D_1D^*\pi$ couplings.  Including the $S$-wave $D_1D^*\pi$ vertex has a certain impact on the triangle diagram mechanism for the $\zc$ peak, since it provides a sizable different background contribution to the $J/\psi\pi$ invariant mass distribution. Moreover, we have used in this work dimensional regularization to render the integrals in the Lippmann-Schwinger equation UV finite. This resolves the issue of employing Gaussian regulators, as in Refs.~\cite{Albaladejo:2015lob,Yang:2020nrt}, which could significantly enhance (weaken) the interaction strength  in the energy region far below (above) the relevant channel threshold. Finally, we have investigated effects from SU(3) light-flavor violations, which are found to be moderate and of the order of 20\%. 

The successful reproduction of the BESIII measured spectra, in both non-strange and strange hidden-charm sectors, strongly supports that the $\zcs$ and $\zc$ are SU(3) flavor partners placed in the same octet multiplet. The best results are obtained when an energy-dependent term in the diagonal $D^{(*)}\bar D_{(s)}^{(*)}$  interaction is included, leading  to resonances (poles above the corresponding open-charm thresholds) to describe these exotic states, though the data are also compatible with the $\zc$ and $\zcs$ as virtual states (poles below the corresponding open-charm thresholds on the real energy axis of the unphysical RS). We have also made predictions for the isovector $Z_{c}^*$ and isodoublet $Z_{cs}^*$, $D^*\bar{D}^*$ and $D^*\bar{D}_{s}^*$ molecules, with $J^{PC}=1^{+-}$ and $J^{P}=1^{+}$, respectively. These states would be HQSS partners of the $Z_{c}$ and $Z_{cs}$. The masses and widths of the  $\zc$, $\zcs$, $Z_c^*$, and $Z_{cs}^*$ resonances are collected in Table~\ref{tab:pole_su3b}.

One important feature of the contributions from triangle diagrams with a TS close to the physical region is the sensitivity to the kinematic variables~\cite{Guo:2019twa}. In the problem under study, the significance of the $Z_{cs}(3885)$ signal at the $e^+e^-$ c.m. 4.681~GeV relative to the other energies can be attributed to this effect. 
At this respect and to better understand the nature of the $Z_c(3900)$, high-statistic  data in a sufficiently large  range of $e^+e^-$ c.m. energies, including not only 4.23 and 4.26~GeV, but also for instance  4.29~GeV, where the TS plays a more important role, and other energies far from 4.29 GeV will be highly valuable. In this way, one should have enough information to map out the relative important on the relevant invariant mass spectra of the TS and the pole contributions. 

\medskip

\begin{acknowledgments}
We would like to thank Rong-Gang Ping for helpful discussions regarding the energy resolution in BESIII measurements.
This work is supported in part by the Spanish Ministry of Science and Innovation (MICINN) (Project PID2020-112777GB-I00), by the EU Horizon
2020 research and innovation programme, STRONG-2020 project, under grant agreement No.~824093, by Generalitat
Valenciana under contract PROMETEO/2020/023, by the National Natural Science Foundation of China (NSFC) and the
Deutsche Forschungsgemeinschaft (DFG) through the funds provided to the Sino-German Collaborative Research Center TRR110 ``Symmetries and the Emergence of Structure in QCD'' (NSFC Grant No. 12070131001, DFG Project-ID 196253076),
by the NSFC under Grants No.~12125507, No.~11835015, No.~12047503, and No.~11961141012, and by the Chinese Academy of Sciences (CAS) under Grants
No. XDB34030000, No.~XDPB15 and No. QYZDB-SSW-SYS013. M.A. is supported by Generalitat Valenciana under Grant No. CIDEGENT/2020/002. J.N. is also supported by the CAS President's International Fellowship Initiative under Grant No.~2020VMA0024.
\end{acknowledgments}

\appendix
\section{Three-point scalar loop function}\label{sec:appen}

The scalar three-point ($P_1P_2P_3$) loop function, see, {\it e.g.}, Eq.~\eqref{eq:3-point}, can be calculated analytically\footnote{Note that, unlike in Refs.~\cite{Albaladejo:2015dsa,Albaladejo:2015lob}, we do not include any form factor in the numerator of the three-point loop function, because of the dimension regularization scheme employed in this work.} in terms of simple elementary functions when all of the three intermediate particles $P_1$, $P_2$ and $P_3$, are treated nonrelativistically. The scalar three-point loop integral for the reaction $A\to B+C$ in Fig.~\ref{fig:3-point} can be expressed as~\cite{Guo:2010ak,Guo:2019twa}
\bea
I(s) &=& i\int\frac{d^4q}{(2\pi)^4} \frac{1}{(q^2-m_1^2+i\epsilon)\left( (P-q)^2-m_2^2+i\epsilon)\right)((q-k)^2-m_3^2+i\epsilon)} \nonumber\\
&\simeq & \frac{\mu_{12}\mu_{23}}{16\pi m_1m_2m_3}\frac{1}{\sqrt{a}}\left[ \arctan \left( \frac{c_2-c_1}{2\sqrt{a(c_1-i\epsilon)}}\right) - \arctan\left( \frac{c_2-c_1-2a}{2\sqrt{a(c_2-a-i\epsilon)}}\right)\right],
\eea
where $c_1=2\mu_{12}b_{12}$, $c_2=2\mu_{23}b_{23}+{ q^2_B}\mu_{23}/m_3$, and $a=(\mu_{23}/m_3)^2q_B^2$, with $\mu_{ij}=m_im_j/(m_i+m_j)$, $b_{12}=m_1+m_2-M$, and $b_{23}=m_2+m_3+E_B-M$. Here we have used the expressions in the rest frame of the initial state, {\it i.e.} $P^\mu = \{ M,\vec{0} \}$, and $k^\mu = \{ E_B, \vec{q}_B \}$, {\it i.e.},
\bea
q_B = \frac{1}{2M}\lambda^\frac12(M^2,m_B^2,s),
\eea
where $s$ is the invariant mass of $C$, {\it i.e.}, $s = (P-k)^2$.

\begin{figure*}[tb!]
\centering
\includegraphics[width=0.4\textwidth]{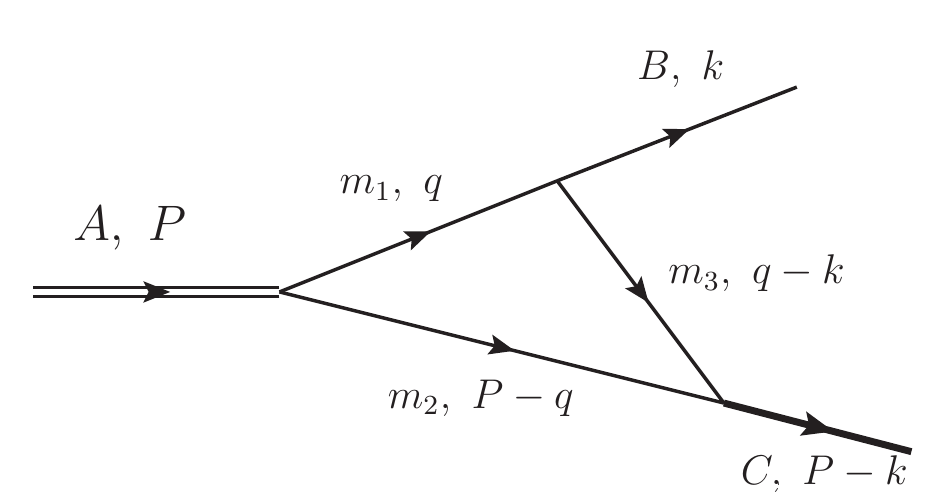}
\caption{A triangle diagram for the reaction $A\to B+C$ through the $P_1P_2P_3$ loop, with $m_i$ the mass of the intermediate particle $P_i$.}
\label{fig:3-point}
\end{figure*}

\bibliography{ZcZcs.bib}
 
\end{document}